\documentclass{article}

\usepackage{arxiv}

\usepackage[utf8]{inputenc} 
\usepackage[T1]{fontenc}    
\usepackage{hyperref}       
\usepackage{url}            
\usepackage{booktabs}       
\usepackage{amsfonts}       
\usepackage{nicefrac}       
\usepackage{microtype}      
\usepackage{graphicx}
\usepackage{amssymb}
\usepackage{amsmath}
\usepackage{lineno}
\usepackage{hyperref}
\usepackage{listings}
\usepackage{subfigure}
\usepackage{comment}
\usepackage{enumitem}
\usepackage[dvipsnames]{xcolor}
\usepackage{algorithm}
\usepackage{algpseudocode}

\title{Solving geophysical inverse problems with measurement-guided diffusion models}

\author{
  Matteo Ravasi \\
  KAUST\\
  Thuwal, Kingdom of Saudi Arabia \\
  \texttt{matteo.ravasi@kaust.edu.sa}}
  
\begin{document}

\chead{Geophysical inverse problems with measurement-guided diffusion models}

\maketitle

\begin{abstract}
  Solving inverse problems with the reverse process of a diffusion model represents an appealing avenue to produce highly realistic, yet diverse solutions from incomplete and possibly noisy measurements, ultimately enabling uncertainty quantification at scale. However, because of the intractable nature of the score function of the likelihood term (i.e., $\nabla_{\mathbf{x}_t} p(\mathbf{y} | \mathbf{x}_t)$), various samplers have been proposed in the literature that use different (more or less accurate) approximations of such a gradient to guide the diffusion process towards solutions that match the observations. In this work, I consider two sampling algorithms recently proposed under the name of Diffusion Posterior Sampling (DPS) and Pseudo-inverse Guided Diffusion Model (PGDM), respectively. In DSP, the guidance term used at each step of the reverse diffusion process is obtained by applying the adjoint of the modeling operator to the residual obtained from a one-step denoising estimate of the solution. On the other hand, PGDM utilizes a pseudo-inverse operator that originates from the fact that the one-step denoised solution is not assumed to be deterministic, rather modeled as a Gaussian distribution. Through an extensive set of numerical examples on two geophysical inverse problems (namely, seismic interpolation and seismic inversion), I show that two key aspects for the success of any measurement-guided diffusion process are: i) our ability to re-parametrize the inverse problem such that the sought after model is bounded between -1 and 1 (a pre-requisite for any diffusion model); ii) the choice of the training dataset used to learn the implicit prior that guides the reverse diffusion process. Numerical examples on synthetic and field datasets reveal that PGDM outperforms DPS in both scenarios at limited additional cost.
\end{abstract}

\section{Introduction}

Many problems in geophysics can be naturally formalized as Bayesian inference latent models, where the target is represented by the posterior distribution $p(\mathbf{x}|\mathbf{y}) \alpha p(\mathbf{y}|\mathbf{x}) p(\mathbf{x})$ given by the observations $\mathbf{y}=g(\mathbf{x}) + \mathbf{n}$ and a prior distribution $p(\mathbf{x})$. Generative models have recently proved to be remarkable priors for Bayesian inference, and several successful applications to geophysical inverse problems can be found in the literature. Likely due to their early popularity in the realm of generative models, Generative Adversarial Networks (GANs) have been first employed in the solution of full-waveform inversion (e.g.,~\cite{Mosser2020}) and history matching (e.g.,~\cite{Mosser2019}) problems. This is accomplished by repeatedly solving the inverse problem at hand for the parameters of latent vector of the generator, producing expressive solutions that align with the physics of the modeling operator (i.e., likelihood function) and the prior expectation as captured by the training dataset. In a similar fashion,~\cite{Ravasi2023a} proposed to use the decoder of a Variational AutoEncoder (VAE) network to parametrize the solution of a variety of inverse problems with applications to seismic processing. By casting the inference process in a variational setting (i.e., using an objective function grounded in the theory Variational Inference -- VI) and solving for the free-parameters of the so-called proposal distribution, one can ultimately generate a possibly infinite number samples from the approximate posterior distribution. Normalizing Flows (NFs) have also been successfully used in a variety of problems to parametrize the proposal distribution in an amortized-VI formulation, with applications ranging from interpolation~\cite{Kumar2021}, to imaging~\cite{Siahkoohi2023}, to full-waveform inversion~\cite{Sun2024}. This approach requires one to train a conditional NF end-to-end assuming access to pairs of training data (e.g., seismic images and reflectivity models); a refinement step that relies on the physical modeling operator can be further employed to mitigate moderate data distribution shifts at inference time. Last but not least, geophysicists have begun exploring diffusion and score-based models as effective tools to implicitly capture the prior knowledge embedded in a representative training set. These models are being applied to address inverse problems such as seismic interpolation~\cite{Ravasi2023b}, seismic demultiple~\cite{Durall2023}, seismic inversion~\cite{Song2024}, and full-waveform inversion~\cite{Wang2023}.

All of the aforementioned approaches rely on a representative training dataset to train either a generative model that later acts as an implicit prior in the solution of an inverse problem or a conditional generative model; unsupervised approaches have been recently proposed as an alternative solution in the absence of representative training datasets. For example, \cite{Elmeliegy2024} used a VAE to parametrize the velocity model in full-waveform inversion and optimize its parameters by means of a physics-based loss (i.e., the classical L2 misfit between observed and modeled data). Similarly,~\cite{Rizzuti2020} and ~\cite{Rizzuti2024} suggested to solve the post-stack seismic inversion problem by parametrizing the sought after acoustic impedance model with a NF, whose parameters are once again optimized using a physics-based loss. Finally,~\cite{Romero2024} followed a similar direction using an Implicit Neural Representation (INR) network to parametrize the acoustic impedance model directly from coordinates of the domain of interest.

Whilst all of the methods mentioned above come with their unique set of opportunities and challenges, diffusion models possibly represent the most appealing choice of generative model to be used as prior in the solution of geophysical inverse problems because of their versatility and ability to produce  samples of extremely high visual quality. However, obtaining exact Bayesian posterior samples for these models remains intractable, necessitating the use of approximations. Current methods for this approximate sampling can be broadly grouped into three categories: the first family of methods trains a conditional diffusion model where the measurements are used as the condition~\cite{Li2021, Saharia2022}; the second family of methods modifies a standard diffusion sampling process by enforcing data consistency at every time step, ensuring that all intermediate samples align with the data measurements~\cite{Song2020, Kadkhodaie2021, Chung2022a, Chung2022b, kawar2022denoising}. The third approach estimates the score function (i.e., the gradient of the log probability density) of the likelihood function, leveraging it to guide each diffusion sampling iteration~\cite{Jalal2021, kawar2021denoising, chung2023diffusion, Song2023}. The latter two approaches are appealing in general contexts as they decouple the training process of the diffusion model with the solution of the inverse problem at hand. I refer the reader to~\cite{Daras2024} for a comprehensive review of these approaches an other approaches proposed to date to solve inverse problems with diffusion models. 

In this work, I assess the performance of two of such sampling algorithms proposed under the name of Diffusion Posterior Sampling (DPS) and Pseudo-inverse Guided Diffusion Model (PGDM), respectively. In DSP, the guidance term used at each step of the reverse diffusion process is obtained by applying the adjoint of the modeling operator to the residual obtained from a one-step denoising estimate of the solution. On the other hand, PGDM utilizes a pseudo-inverse operator that originates from the fact that the one-step denoised solution is not assumed to be deterministic, rather modeled as a Gaussian distribution. Moreover, because of the similarity between DPS and another sampling method called Manifold Constrained Gradient (MCG), where the latter performs an additional projection onto the measurement subspace after each diffusion update step, I also evaluate MCG in the case of noise-free problems. Through an extensive set of numerical examples on two geophysical inverse problems (namely, seismic interpolation and seismic inversion), I show that two key aspects for the success of any measurement-guided diffusion process are: i) our ability to re-parametrize the inverse problem such that the sought after model is bounded between -1 and 1 (a pre-requisite for any diffusion model); ii) the choice of the training dataset used to learn the implicit prior that guides the reverse diffusion process. Numerical examples on synthetic and field datasets reveal that PGDM outperforms DPS in both scenarios at limited additional cost.

\section{Theory}

Diffusion models are a form of latent variable models used to learn a complex distribution $p(\mathbf{x}_0)$ directly from samples of such a distribution, $X=\{\mathbf{x}^1_0, \mathbf{x}^2_0, \mathbf{x}^{n_s}_0\}$, where $n_s$ is the number of available training samples. More specifically, these models are composed of 2 processes:

\begin{itemize}
\item a fixed (or predefined) forward diffusion process $p$ that gradually adds Gaussian noise to an input $\mathbf{x}_0 \sim p(\mathbf{x}_0)$, until it is transformed into pure noise, $\mathbf{x}_T \sim \mathcal{N}(\mathbf{x}_T; \mathbf{0}, \mathbf{I})$ (where $T$ represents the number of steps of the diffusion process);
\item a learned reverse denoising diffusion process $q_\theta$, where a neural network is trained to gradually denoise an input composed of pure noise, until the forward process is completely reversed to produce an output $\hat{\mathbf{x}}_0$ that belongs to the distribution of interest $p(\mathbf{x}_0)$.
\end{itemize}
In the literature, diffusion models have been described both in a discretized (i.e., via Markov chains) and continuous form (i.e., via stochastic differential equations -- SDEs). For the sake of simplicity, in this work I consider the discrete case. Moreover, there exist two families of diffusion processes, namely Variance Exploding (VE) and Variance Preserving (VP) processes, which differ in how the signal is scaled throughout the diffusion process. Again, I only consider the VP case here, although the VE case is equivalent to the VP case up to a time-dependent scaling factor~\cite{kawar2022denoising}, so any sampling algorithm for the VP case can be adapted to the VE case as well.

\subsection{Forward process} 
The forward process of a VP diffusion model can described by the following transition of a Markov chain:
\begin{equation}
p(\mathbf{x}_t | \mathbf{x}_{t-1}) = \mathcal{N}(\mathbf{x}_t; \sqrt{1-\beta_t}\mathbf{x}_{t-1}, \beta_t \mathbf{I}),
\label{eq:diffforward}
\end{equation}
where $0<\beta_1<\beta_2<...<\beta_t<...<\beta_T<1$ represents the so-called variance schedule. In other words, each new (slightly noisier) sample of the diffusion process is drawn from a Gaussian distribution whose mean corresponds to a scaled version of the previous sample of the chain and whose variance is fully defined by the chosen variance schedule. Note that by scaling $\mathbf{x}_{t-1}$ with $\sqrt{1-\beta_t}$ at each step of the chain, the overall process is guaranteed to have a bounded variance (equal to 1 if $p(\mathbf{x}_0)$ has unitary variance). This corresponds to a VP-SDE in the continuous case. On the other hand, if the scaling $\sqrt{1-\beta_t}$ is omitted from the update rule, the variance of the overall process becomes unbounded, leading to a VE-SDE in the continuous case.

A direct consequence of the constructed $p$ process is that one can sample $\mathbf{x}_t$ at any noise level $t$ directly from $\mathbf{x}_0$ (and without having to run through the entire chain from $0$ to $t$). More precisely, one can write:
\begin{equation}
p(\mathbf{x}_t | \mathbf{x}_0) = \mathcal{N}(\mathbf{x}_t; \sqrt{\bar{\alpha}_t}\mathbf{x}_0, (1-\bar{\alpha}_t) \mathbf{I}),
\label{eq:diffforwardnsteps}
\end{equation}
where $\alpha_t=1-\beta_t$ and $\bar{\alpha}_t=\prod_{s=1}^t \alpha_s$. This property is particularly advantageous during the training process, as one can create samples with different noise levels directly from any input sample $\mathbf{x}_0$ as follows
\begin{equation}
\mathbf{x}_t = \sqrt{\bar{\alpha}_t}\mathbf{x}_0 + \sqrt{1-\bar{\alpha}_t} \boldsymbol \epsilon,
\label{eq:diffforwardnsteps1}
\end{equation}
and then optimize random terms of the loss function as discussed later in more details.

\subsection{Reverse process}
The goal of the reverse process is to run through the chain in reverse order. This could be achieved in a similar manner to the forward process if the reverse transition $p(\mathbf{x}_{t-1} | \mathbf{x}_t)$ was known a-priori. However since it is intractable to directly compute (or sample from) such a probability, a neural network is instead leveraged to approximate it. As such, I refer to it as $q_\theta(\mathbf{x}_{t-1} | \mathbf{x}_t)$ with $\theta$ representing the learnable parameters of the network. More precisely, by assuming that the reverse process is also Gaussian, with $q(\mathbf{x}_T) \sim \mathcal{N}(\mathbf{x}_T; \mathbf{0}, \mathbf{I})$, this distribution can be defined as follows:
\begin{equation}
q_\theta(\mathbf{x}_{t-1} | \mathbf{x}_t) = \mathcal{N}(\mathbf{x}_{t-1}; \boldsymbol \mu_\theta(\mathbf{x}_t, t), \boldsymbol \Sigma_\theta(\mathbf{x}_t, t)),
\label{eq:diffreverse}
\end{equation}
where both the mean and covariance matrix are conditioned on the step $t$. In practice, the authors of one of the most effective diffusion models (namely, the Denoising Diffusion Probabilistic Model -- DDPM --~\cite{ho2020denoising}) decide to learn only the mean and fix the covariance matrix to $\Sigma_\theta(\mathbf{x}_t, t)=\sigma_t^2 \mathbf{I}$, where $\sigma_t=\beta_t$ or $\sigma_t=(1-\bar{\alpha}_{t-1}) / (1-\bar{\alpha}_t) \beta_t$. Alternatively, the covariance matrix can also be learned as shown in~\cite{Nichol2021}.

\subsection{Training a diffusion model}
Training a diffusion model encompasses estimating the mean (and possibly covariance) of the reverse transition $q_\theta(\mathbf{x}_{t-1} | \mathbf{x}_t)$. Starting from the variational bound on the negative log-likelihood (see~\cite{ho2020denoising} for a detailed derivation):
\begin{equation} 
    \label{eq:varboundlike}
    \mathcal{L} := \mathbb{E}_{\mathbf{x}_0 \sim p(\mathbf{x}_0)} \left[ - log q_\theta(\mathbf{x}_0))\right] \le \mathbb{E}_{\mathbf{x}_0 \sim p(\mathbf{x}_0)} \left[ - log q_\theta(\mathbf{x}_T) - \sum_{t\ge 1} log \frac{q_\theta(\mathbf{x}_{t-1} | \mathbf{x}_t)}{p(\mathbf{x}_t | \mathbf{x}_{t-1})} \right],
\end{equation}
and by taking advantage of equation \ref{eq:diffforwardnsteps}, training can be performed in an efficient manner by optimizing random terms of $\mathcal{L}$ with a stochastic gradient descent algorithm. More specifically, for the case of fixed covariance, the loss function becomes:
\begin{equation} 
    \label{eq:lossmu}
    \mathcal{L} = \mathbb{E}_{\mathbf{x}_0 \sim p(\mathbf{x}_0),  t \sim \mathcal{U}(1, T)} \left[ \frac{1}{2\sigma_t^2}||\tilde{\boldsymbol \mu}_t(\mathbf{x}_t, \mathbf{x}_0) - \boldsymbol \mu_\theta(\mathbf{x}_t, t)||_2^2
     \right],
\end{equation}
where $\tilde{\boldsymbol \mu}_t(\mathbf{x}_t, \mathbf{x}_0)$ is the mean of the reverse transition probability further conditioned on $\mathbf{x}_0$:
\begin{equation} 
    \label{eq:pcond0}
    p(\mathbf{x}_{t-1} | \mathbf{x}_t, \mathbf{x}_0) \sim \mathcal{N}(\mathbf{x}_{t-1};  \tilde{\boldsymbol \mu}_t(\mathbf{x}_t, \mathbf{x}_0), \tilde{\beta}_t \mathbf{I}), 
\end{equation}
and equals
\begin{equation}
\tilde{\boldsymbol \mu}_t(\mathbf{x}_t, \mathbf{x}_0) = \frac{\sqrt{\bar{\alpha}_{t-1}} \beta_t}{1-\bar{\alpha}_t}\mathbf{x}_0+\frac{\sqrt{\alpha_t}(1-\bar{\alpha}_{t-1})}{1-\bar{\alpha}_t}\mathbf{x}_t \qquad \tilde{\beta}_t =\frac{1-\bar{\alpha}_{t-1}}{1-\bar{\alpha}_t}\beta_t.
\end{equation} 
However, given an alternative parametrization for $\tilde{\boldsymbol \mu}$, the loss function can be simplified to:
\begin{equation} 
    \label{eq:varboundlike1}
    L = \mathbb{E}_{\mathbf{x}_0 \sim p(\mathbf{x}_0), t \sim \mathcal{U}(1, T), \boldsymbol \epsilon \sim \mathcal{N}(\boldsymbol \epsilon; \mathbf{0}, \mathbf{I})} \left[ \frac{\beta_t^2}{2 \sigma_t^2 \alpha_t (1-\bar{\alpha}_t)} || \boldsymbol \epsilon - \boldsymbol \epsilon_\theta (\mathbf{x}_t=\sqrt{\bar{\alpha}_t}\mathbf{x}_0 + \sqrt{1-\bar{\alpha}_t} \boldsymbol \epsilon, t) ||_2^2 \right],
\end{equation}
where one can see that the learning process of a diffusion model simply corresponds to predicting the unscaled noise added to $\mathbf{x}_0$ at each step $t$, where $\mathbf{x}_t$ is conveniently computed using equation~\ref{eq:diffforwardnsteps1}. This closely resembles the denoising score matching objective function when trained over multiple noise scales indexed by $t$. In practice,~\cite{ho2020denoising} found that discarding the scaling term leads to a more efficient training process.

\subsection{Diffusion models as priors for inverse problems}
Let us consider an inverse problem of the form:
\begin{equation} 
    \label{eq:problem}
    \mathbf{d} = g(\mathbf{x}_0) + \mathbf{w},
\end{equation}
where the model $\mathbf{x}_0$ and the observed data $\mathbf{d}$ are connected via a nonlinear operator $g$ (or a linear operator $\mathbf{G}$) and the observed data is contaminated by noise $\mathbf{w}$. Here, both model and noise are assumed here to be realizations of given probability distributions, namely $\mathbf{x}_0\sim p_x(\mathbf{x}_0)$ and $\mathbf{w} \sim p_w(\mathbf{w})$. More specifically, I represent the noise with an i.i.d. Gaussian distribution with zero mean and known standard deviation, i.e. $ \mathbf{w} \sim \mathcal{N}(\mathbf{w}; \mathbf{0}, \sigma_w^2 \mathbf{I})$. The inverse problem in equation \ref{eq:problem} can be then casted in a Bayesian form with the help of Bayes' Theorem, providing a closed-form solution for the posterior distribution $p(\mathbf{x}_0|\mathbf{d})$ (up to a normalizing constant $Z$):
\begin{equation}
	\label{eq:bayes}
    p(\mathbf{x}_0|\mathbf{d}) = \frac{1}{Z}p_w(\mathbf{d}|\mathbf{x}_0)p_x(\mathbf{x}_0),
\end{equation}
where $p_w(\mathbf{d}|\mathbf{x}_0)$ is the so-called likelihood function, whilst $p_x(\mathbf{x}_0)$ is the prior distribution, which in this work I aim to learn by means of a diffusion model.

In theory, it is straightforward to extend the reverse diffusion process to include the effect of the likelihood function (and so that of the data and the modeling operator for a specific inverse problem) alongside that of the prior. As discussed in~\cite{chung2023diffusion} and~\cite{Song2023}, this can be achieved by starting from the SDE of the reverse process:
\begin{equation}
	\label{eq:sdereverse}
    d\mathbf{x} = \left[ -\frac{\beta(t)}{2} \mathbf{x} - \beta(t)\nabla_{\mathbf{x}_t} log p_t(\mathbf{x}_t)\right] dt + \sqrt{\beta(t)}d\mathbf{w},
\end{equation}
where $dt$ corresponds to time running backward and $d\mathbf{w}$ is the standard Wiener process running backward, and modifying the term inside the square brackets (usually referred to as the drift term) to include also the effect of the likelihood function:
\begin{equation}
	\label{eq:sdereverseposterior}
    d\mathbf{x} = \left[ -\frac{\beta(t)}{2} \mathbf{x} - \beta(t) (\nabla_{\mathbf{x}_t} log p_t(\mathbf{x}_t) + \nabla_{\mathbf{x}_t} log p_{w,t}(\mathbf{d} | \mathbf{x}_t))\right] dt + \sqrt{\beta(t)}\mathbf{w}
\end{equation}
However, whilst the score function of the prior distribution can be easily computed from a pre-trained, problem-agnostic diffusion model (i.e., $\nabla_{\mathbf{x}_t} log p_t(\mathbf{x}_t) \approx \boldsymbol\epsilon_{\bar{\theta}}(\mathbf{x}_t, t)$), the score function of the likelihood term ($\nabla_{\mathbf{x}_t} log p_{w,t}(\mathbf{d} | \mathbf{x}_t)$) is intractable to compute given that both $\textbf{x}_t$ and $\textbf{d}$ depend on $\textbf{x}_0$, but they are independent from each other. Initial implementations of measurement-guided reverse diffusion models, aimed at solving inverse problems, assumed that
$p_{w,t}(\mathbf{d} | \mathbf{x}_t)\approx p_w(\mathbf{d} | \mathbf{x}_0)$, and therefore computed the score function of the likelihood as:
\begin{equation}
	\label{eq:ald}
    \nabla_{\mathbf{x}_t} log p_{w,t}(\mathbf{d} | \mathbf{x}_t) \approx - \frac{1}{2\sigma_w^2}\nabla_{\mathbf{x}_t} ||\textbf{d}-g(\hat{\mathbf{x}}_t)||_2^2,
\end{equation}
Although initially proposed in the context of MRI imaging, this approach can be easily applied to any linear or nonlinear inverse problem, and is usually referred in the literature as Score-Based Annealed Langevin Dynamics (Score-ALD --~\cite{Jalal2021}). However, this method has been shown to under-perform in practical scenarios due to the strong assumption that $\mathbf{x}_t\approx \mathbf{x}_0$; I will therefore not consider it in the numerical examples.

Alternatively, one can be partially circumvented the issue of having an intractable score function for the likelihood term by marginalizing out $\textbf{x}_0$ as follows:
\begin{equation}
	\label{eq:sdereverseposteriorintegral}
    p_{w,t}(\mathbf{d} | \mathbf{x}_t) = \int_{\textbf{x}_0} p_w(\mathbf{d} | \mathbf{x}_0) p_t(\textbf{x}_0 | \mathbf{x}_t) d \textbf{x}_0.
\end{equation}
and by noting that the likelihood function $p_w(\mathbf{d} | \mathbf{x}_0)$ is tractable (and with closed-form solution in case of Gaussian noise), whilst an approximation of $p_t(\textbf{x}_0 | \mathbf{x}_t)$ can be obtained from equation \ref{eq:diffforwardnsteps1}. More specifically, one can reconstruct the original sample from the sample at time step $t$ as follows (i.e., one-step denoising):
\begin{equation} 
	\label{eq:onestepdenoising}
   \mathbf{x}_0 \approx  \hat{\mathbf{x}}_0 = \frac{1}{\sqrt{\bar{\alpha}_t}}\left(\mathbf{x}_t- \sqrt{1-\bar{\alpha}_t}\boldsymbol \epsilon_\theta(\mathbf{x}_t, t)\right).
\end{equation}
\cite{chung2023diffusion} derive their reverse sampling process using the following approximation: $ p_{w,t}(\mathbf{d} |\mathbf{x}_t) \approx p_w(\mathbf{d} | \hat{\mathbf{x}}_0)$. For the case of Gaussian noise, the following score function is added to the reverse step:
\begin{equation}
	\label{eq:dpsscore}
   \nabla_{\mathbf{x}_t} log p_{w,t}(\mathbf{d} | \mathbf{x}_t) \approx - \frac{1}{2\sigma_w^2}\nabla_{\mathbf{x}_t} ||\textbf{d}-g(\hat{\mathbf{x}}_0)||_2^2,
\end{equation}
which for the linear case becomes
\begin{equation}
	\label{eq:dpsscorelin}
   \nabla_{\mathbf{x}_t} log p_{w,t}(\mathbf{d} | \mathbf{x}_t) \approx \frac{1}{\sigma_w^2}(\nabla_{\mathbf{x}_t}\hat{\mathbf{x}}_0)^T \textbf{G}^T (\textbf{d}-\textbf{G}\hat{\mathbf{x}}_0).
\end{equation}
where the superscript $T$ is used to represent the adjoint of the operator $\textbf{G}$, and $\nabla_{\mathbf{x}_t} \hat{\mathbf{x}}_0$ is the Jacobian of equation~\ref{eq:onestepdenoising} (which contains the Jacobian of the neural network used to predict the unscaled noise). I refer to this as the Diffusion Posterior Sampling (DPS) algorithm. 

The authors of the DPS algorithm have almost concurrently proposed a variant of this method that usually works well with noise-free data (whilst being sub-par to DPS in the case of noisy data). This approach, named the Manifold Constrained Gradient (MCG), performs the same update step of DPS; however, it introduces an additional projection step onto the measurement subspace of the form of $\mathbf{x}_{t-1}=(\mathbf{I}- \mathbf{G}^T\mathbf{G})\mathbf{x}_{t-1} + \mathbf{G}^T \mathbf{d}'$ where $\mathbf{d}'=\sqrt{\bar{\alpha}_t}\mathbf{d} + \sqrt{1-\bar{\alpha}_t}\mathbf{z}$ where $\mathbf{z} \sim \mathcal{N}(\mathbf{0},\mathbf{I})$. In other words, since the current estimate  $\mathbf{x}_{t-1}$ is still affected by the noise of the diffusion process, the data that is used for the projection is also contaminated by the same amount of noise.

Alternatively,~\cite{Song2023} show that $p_t(\mathbf{x}_0 |\mathbf{x}_t)$ can be more accurately approximated by a Gaussian distribution (instead of a point estimate) with probability density function equal to
\begin{equation}
	\label{eq:onestepdenoisingprob1}
   p_t(\mathbf{x}_0 |\mathbf{x}_t) \approx  \mathcal{N}(\hat{\mathbf{x}}_0, r_t^2 \mathbf{I}).
\end{equation}
In the linear case, this leads to the following expression for the likelihood term
\begin{equation}
	\label{eq:likepgdm}
	p_{w,t}(\mathbf{d} |\mathbf{x}_t) \sim \mathcal{N}(\mathbf{G}\hat{\mathbf{x}}_0, r_t^2\mathbf{G}\mathbf{G}^T+\sigma_w^2\textbf{I}),
\end{equation}
whose score function is
\begin{equation}
	\label{eq:pgdmcore}
   \nabla_{\mathbf{x}_t} log p_{w,t}(\mathbf{d} | \mathbf{x}_t) \approx (\nabla_{\mathbf{x}_t} \hat{\mathbf{x}}_0)^T \mathbf{G^T}(r_t^2\textbf{G}\textbf{G}^T+\sigma_w^2\mathbf{I})^{-1} (\textbf{d}-\mathbf{G}\hat{\mathbf{x}}_0) 
\end{equation}
I refer to this as the Pseudo-inverse Guided Diffusion Model (PGDM) algorithm. In practice the pseudo-inverse is never formed explicitly; instead the following problem is solved using an iterative algorithm (e.g., conjugate gradient) for a given number of iterations: $ (r_t^2\textbf{G}\textbf{G}^T+\sigma_w^2\mathbf{I}) \textbf{g} = \textbf{d}-\mathbf{G}\hat{\mathbf{x}}_0$, and $\textbf{g}$ is then multiplied to the remaining terms in equation~\ref{eq:pgdmcore}. Finally, if one considers the case where measurement noise is absent (i.e., $\sigma_w^2=0$), equations~\ref{eq:pgdmcore} simplifies to:
\begin{equation}
	\label{eq:pgdmnoisefree}
   \nabla_{\mathbf{x}_t} log p_{w,t}(\mathbf{d} | \mathbf{x}_t) \approx \frac{1}{r_t^2}(\nabla_{\mathbf{x}_t}\hat{\mathbf{x}}_0)^T \textbf{G}^\dagger (\textbf{d}-\textbf{G}\hat{\mathbf{x}}_0),
\end{equation}
with $\mathbf{G}^\dagger=\mathbf{G}^T(\mathbf{G}\mathbf{G}^T)^{-1}$. Note the close similarity with equation \ref{eq:dpsscorelin} where the pseudo-inverse replaces the transpose of the operator (i.e., a spectral correction is applied).~\cite{Song2023} state that the former approximation can be acceptable in the presence of modeling operators with small conditioning number, whilst it may affect the convergence of the sampling process for ill-conditioned modeling operators. 

\subsection{Algorithms}
In the following, I summarize the training procedure and various measurement-guided reverse processes used in this work. For the training algorithm, it is important to notice that whilst the training loop is written for a single sample, this is generally carried out using batches of samples and randomly selected $t$ steps. In the measurement-guided sampling algorithms instead, $\tilde{\mathbf{x}}_0$ is intended to be a problem-dependent starting guess (which will be detailed later for each problem of interest), whilst $\mathbf{g}_t$ and $\eta_t$ are the guidance vector and step size, respectively. Depending on the choice of the sampling algorithm, and considering the linear case, they correspond to:
\begin{itemize}
\item Score-ALD: $\mathbf{g}_t = - \mathbf{G}^T (\mathbf{d} - \mathbf{G x}_t)$, $\eta_t=\frac{1}{\sigma^2+\gamma_t}$ where $\gamma_t$ is a step-dependent hyperparameter.
\item DPS/MCG: $\mathbf{g}_t = - (\nabla_{\mathbf{x}_t}\hat{\mathbf{x}}_0)^T \textbf{G}^T (\textbf{d}-\textbf{G}\hat{\mathbf{x}}_0)$, $\eta_t=\frac{\eta'}{||\textbf{d}-\textbf{G}\hat{\mathbf{x}}_0||_2}$ where $\eta'$ is an hyperparameter.
\item PGDM with orthonormal operator (i.e., $\textbf{G}\textbf{G}^T = \textbf{I}$): $\mathbf{g}_t = - (1 + \sigma_w^2 / r_t^2)^{-1} (\nabla_{\mathbf{x}_t}\hat{\mathbf{x}}_0)^T \textbf{G}^T (\textbf{d}-\textbf{G}\hat{\mathbf{x}}_0)$ where $r_t=\sqrt{\sigma_t^2/(1+\sigma_t^2)}$ and $\eta_t=\sqrt{\bar{\alpha_t}}$. Note that in the case of $\sigma_w^2=0$, this reconducts to the DPS update albeit with different scaling $\eta_t$;
\item PGDM: $\mathbf{g}_t = - (\nabla_{\mathbf{x}_t}\hat{\mathbf{x}}_0)^T \textbf{G}^ T (\textbf{G} \mathbf{G}^T + \sigma_w^2 / r_t^2 \mathbf{I})^{-1}  (\textbf{d}-\textbf{G}\hat{\mathbf{x}}_0)$ where $r_t$ and $\eta_t$ are the same as those defined in the previous case.
\end{itemize}

\begin{algorithm}
\caption{Training}\label{alg:training}
\begin{algorithmic}
\Require $\mathbf{x}_0, \{\beta_t, t=1,...,T\}$
\Ensure Trained $\epsilon_\theta(\cdot, t)$
\Repeat
\State $\mathbf{x}_0 \sim p(\mathbf{x}_0)$
\State $t \sim \mathcal{U}(1, T)$
\State $\boldsymbol \epsilon \sim \mathcal{N}(\mathbf{0},\mathbf{I})$
\State $\mathcal{L} = ||\boldsymbol \epsilon - \boldsymbol \epsilon_\theta(\sqrt{\bar{\alpha}_t}\mathbf{x}_0 + \sqrt{1-\bar{\alpha}_t} \boldsymbol \epsilon, t)||_2^2$
\State $\theta \gets \theta - \gamma \nabla_\theta \mathcal{L}$
\Until{converged}
\end{algorithmic}
\end{algorithm}


\begin{algorithm}
\caption{Unconditional (DDPM-style) sampling with N steps}
\label{alg:unc}
\begin{algorithmic}
\Require Trained $\epsilon_\theta(\cdot, t), \{\beta_t, t=1,...,T\}, \{V=(v_1,... ,v_N), v_1=1, v_N=T\}$
\Ensure $\hat{\mathbf{x}}_0$
\State $\mathbf{x}_T \sim \mathcal{N}(\mathbf{0},\mathbf{I})$
\For{$i=N,..., 1$}
\State $t=v _i, s=v_{i-1}$                 
\State $\mathbf{z} \sim \mathcal{N}(\mathbf{0},\mathbf{I})$ if $t>1$ else $\mathbf{z}=\mathbf{0}$
\State $\hat{\mathbf{x}}_0 = \frac{1}{\sqrt{\bar{\alpha}_t}}\left(\mathbf{x}_t -\sqrt{1-\bar{\alpha}_t} \boldsymbol \epsilon_\theta(\mathbf{x}_t, t)\right)$
\State $\mathbf{x}_s = \frac{\sqrt{\alpha_t}(1-\bar{\alpha}_s)}{1-\bar{\alpha}_t}\mathbf{x}_t + \frac{\sqrt{\bar{\alpha}_s}\beta_t}{1-\bar{\alpha}_t}\hat{\mathbf{x}}_0 + \sigma_t \mathbf{z}$
\EndFor
\end{algorithmic}
\end{algorithm}

\begin{algorithm}
\caption{Measurement-guided (DDPM-style) sampling warm-started at $T_s$ with N steps}
\label{alg:cond}
\begin{algorithmic}
\Require Trained $\epsilon_\theta(\cdot, t), \{\beta_t, t=1,...,T\}, \{V=(v_1,... ,v_N), v_1=1, v_N=T_s\}, \mathbf{d}, \mathbf{G}, \tilde{\mathbf{x}}_0$
\Ensure $\hat{\mathbf{x}}_0$
\State $\mathbf{x}_{T_s} = \mathcal{N}(\mathbf{x}_{T_s}; \sqrt{\bar{\alpha}_{T_s}}\tilde{\mathbf{x}}_0 , (1-\bar{\alpha}_{T_s}) \mathbf{I})$
\For{$t=N,..., 1$}    
\State $t=v _i, s=v_{i-1}$                                 
\State $\mathbf{z} \sim \mathcal{N}(\mathbf{0},\mathbf{I})$ if $t>1$ else $\mathbf{z}=\mathbf{0}$
\State $\hat{\mathbf{x}}_0 = \frac{1}{\sqrt{\bar{\alpha}_t}}\left(\mathbf{x}_t -\sqrt{1-\bar{\alpha}_t} \boldsymbol \epsilon_\theta(\mathbf{x}_t, t)\right)$
\State $\mathbf{g}_t = ...$ (see below)
\State $\mathbf{x}_s = \frac{\sqrt{\alpha_t}(1-\bar{\alpha}_s)}{1-\bar{\alpha}_t}\mathbf{x}_t + \frac{\sqrt{\bar{\alpha}_s}\beta_t}{1-\bar{\alpha}_t}\hat{\mathbf{x}}_0 + \sigma_t \mathbf{z} - \eta_t \mathbf{g}_t$
\State (MCG only: $\mathbf{x}_s=(\mathbf{I} - \mathbf{G}^T\mathbf{G})\mathbf{x}_s + \mathbf{G}^T(\sqrt{\bar{\alpha}_t}\mathbf{d} + \sqrt{1-\bar{\alpha}_t}\mathbf{z}') \quad \mathbf{z}' \sim \mathcal{N}(\mathbf{0},\mathbf{I})$)
\EndFor
\end{algorithmic}
\end{algorithm}

\subsection{Adapting geophysical inverse problems to guided diffusion models}
In the following, I briefly recap the theory of the two geophysical inverse problems that will be considered in the Numerical example section, namely seismic interpolation and post-stack inversion. More importantly, I present a number of modifications that are required to ensure that the model that one wishes to invert for (i.e., $\mathbf{x}_0$) is bounded between -1 and 1 as required by the theory of diffusion models.

\subsubsection{Seismic interpolation}
Seismic interpolation is perhaps the simplest of inverse problems one may ever encounter. This task, usually referred in computer vision as inpainting, simply aims to fill gaps in seismic data by leveraging the information contained in nearby available traces. As such, the forward problem can written as
\begin{equation} 
    \label{eq:interp}
    \mathbf{d} = \mathbf{R}\mathbf{x}_0,
\end{equation}
where $\mathbf{R}$ is a restriction operator that extracts the available traces from $\mathbf{x}_0$. It should be noted that this is a very specific instance of equation \ref{eq:problem}, since I assume the absence of noise in the data. At this stage of processing, all of the information contained in the available traces is commonly considered to be signal; an interpolation algorithm is therefore tasked to reconstruct missing traces whilst \textit{fully} honoring the available ones. Because of this specific property of the modeling operator, the posterior distribution can be very easily computed in the absence of any prior: more specifically, since the observed data does not provide any information of what the solution should look like in the missing traces, the posterior standard deviation in such traces should be equal to infinity. When prior knowledge is introduced to the problem, one expects an overall reduction in standard deviation, with more variability (i.e., higher uncertainty) remaining in the larger gaps and less variability in the smaller gaps. Finally, given the nature of the modeling operator (i.e., $\mathbf{R}\mathbf{R}^T=\mathbf{I}$), one can observe that the PGDM method reduces to the DPS method with simply a different scaling factor $\eta_t$. 

As far as the requirement of having $\mathbf{x}_0$ bounded between -1 and 1, a simple two steps procedure is carried out for the problem of seismic interpolation. First, since the amplitudes of a seismic shot/receiver gather decay with time, a time-dependent scaling factor is applied to the data to balance amplitudes across the entire data. This step is important to ensure a similar signal-to-noise ratio over the entire data when adding noise during the diffusion process. Whilst various approaches could be taken to achieve this goal, I apply a time-space varying weighting factor obtained from the reciprocal of the smoothed envelope of the data, further normalized between 0 and 1. Second, since the modeling operator does not create any mixing between traces (or even between samples in the model vector), by simply normalizing the data between -1 ans 1 one can ensures that the sought after model is also correctly normalized. This is performed as follows:
\begin{equation} 
    \label{eq:interpnorm}
    \mathbf{d}_{norm} = 2 \frac{\mathbf{d} - d_{min}}{d_{max}-d_{min}} - 1,
\end{equation}
where $d_{min}$ and $d_{max}$ are the minimum and maximum values across the entire data. At the end of the reverse diffusion process, the original amplitudes are restored by first applying the inverse of the normalization step:
\begin{equation} 
    \label{eq:interpinvnorm}
    \mathbf{x}_{0} =  \frac{d_{max}-d_{min}}{2}(\mathbf{x}_{0,norm} + 1) + d_{min},
\end{equation}
which is further multiplied by the inverse of the time-dependent scaling factor.

\subsubsection{Post-stack seismic inversion}
Post-stack seismic inversion refers to the process of retrieving an acoustic impedance model from post-stack seismic data. These two quantities can be directly linked to each other via the so-called convolutional model~\cite{Goupillaud1961}:
\begin{equation}
    d(t) \approx  \frac{1}{2} \tilde{w}(t) \ast \frac{d}{dt}\ln(AI(t)),
\label{eq:convmodel}
\end{equation}
where $AI$ is the P-wave (or acoustic) impedance, $\tilde{w}=0.5 w(t)$ is the scaled seismic wavelet, and $\ast$ represents the convolution operator. A discretized version of equation \ref{eq:convmodel} can be written in a compact matrix-vector notation as follows
\begin{equation}
    \mathbf{d} = \mathbf{WDm}
\label{eq:post-stack-matrix}
\end{equation}
where $\mathbf{W}$ represents the convolution operator with a given wavelet (i.e., $0.5  \tilde{w}(t)$), $\mathbf{D}$ is a discretized version of the first-order derivative operator, and $\mathbf{m} = ln(\mathbf{AI})$. I define $\mathbf{G} := \mathbf{WD}$ to denote the post-stack seismic operator.

Opposite to seismic waveforms, geological models present acoustic impedance values that are usually independent of the depth or spatial axes (although a slowly increasing trend is usually observed along the depth axis). As such, no spatially-dependent weighting is required in this case. However, one must take special care when normalizing the model from its natural dynamic range (i.e., $AI_{min}/AI_{max}$) to the required one (i.e., $-1/1$). By assuming knowledge of the expect minimum and maximum values (for example, from available well logs), one can write
\begin{equation}
    \ln(AI(t))_{norm} = 2 * \frac{\ln(AI(t)) - \ln(AI(t))_{min}}{\ln(AI(t))_{max} - \ln(AI(t))_{min}} - 1 = a \ln(AI(t))- b
\label{eq:post-stack-scaling}
\end{equation}
where $a=2/(\ln(AI(t))_{max} - \ln(AI(t))_{min})$ and $b=2 \ln(AI(t))_{min} / ({\ln(AI(t))_{max} - \ln(AI(t))_{min}}) +1$. Because of the linearity of the problem, equation~\ref{eq:convmodel} can be equivalently written as:
\begin{equation}
 a d(t) -  \tilde{w}(t) \ast \frac{d}{dt}b = \tilde{w}(t) \ast \frac{d}{dt}[a \ln(AI(t)) - b],
\label{eq:post-stack-rescaledproblem}
\end{equation}
where the second term on left-hand side goes to zero as the time derivative of a constant is zero. I can therefore solve the following problem:
\begin{equation}
 \tilde{d}(t)= \tilde{w}(t) \ast \frac{d}{dt}\ln(AI(t))_{norm},
\label{eq:post-stack-rescaledproblem1}
\end{equation}
where $\tilde{d(t)} =  a d(t)$. Note that, whilst the above equations are written in a continuous form, one effectively solves the following discretized problem:
\begin{equation}
    \tilde{\mathbf{d}} = \mathbf{Gm}_{norm}
\label{eq:post-stack-matrixscaled}
\end{equation}
and finally rescale the estimated $\ln(AI(t))_{norm}$ back to its original dynamic range. By doing so, the acoustic impedance model to be estimated via a guided reverse diffusion process is ensured to bounded between -1 and 1.

\section{Numerical Examples}
In this section, I present a wide range of numerical examples that entail training diffusion models from scratch on carefully designed training datasets and subsequently deploying them within one of the measurement-guided reverse diffusion processes described above. Note that, since the guidance is effectively provided by the gradient of the likelihood term in the Bayes' theorem (which contain information from both the observed data and modeling operator -- i.e., the physics of the problem), one can solve inverse problems through the reverse process of a diffusion model.

The examples are divided into two main parts: first, I consider the problem of seismic interpolation (working directly with wavefields); second, I approach the problem of post-stack seismic inversion, where wavefields are converted into (blocky) geological models. In both cases, the implicit prior of the solution is captured by a diffusion model previously trained on a representative training set.

\subsection{Training parameters}
The different diffusion models used in this work have been trained in the same way (independently on the downstream inverse problem they will be used for) based on the following parameters:
\begin{itemize}
    \item The diffusion process is composed of 1000 steps and uses either a linear or cosine noise scheduler~\cite{nichol2021improved};
    \item A UNet-like architecture is used for the denoiser. The network is composed of four levels, each containing 2 ResNet layers with two sets of group normalization and convolutional blocks followed by a SiLU activation function. The number of channels across the four levels is increased from 64 to 256 (doubling at every level). For the last two levels, an additional attention block is added at the start of the first ResNet layer. Finally, sinusoidal position embedding is used to inform the network about the particular time step (noise level) is operating at: this information is added after each convolutional block of the ResNet layers by simply shifting the output of the convolutional layer by the positionally encoded time step. Finally, downsampling is performed using an additional convolutional layer with stride of 2, whilst upsampling is performed using nearest neighbor interpolation followed by a convolutional layer.
    \item Training is carried out for 100 epochs using the Adam optimizer with a learning rate of $4e^{-4}$ and a batch size of 64.
\end{itemize}

\subsection{Seismic Interpolation}
In this first example, I consider a synthetic dataset modeled from a subsurface velocity model that mimics the Volve field (see \cite{Ravasi2022} for more details on the data creation process). Next, a similar procedure is applied to a 2D line of the Volve field dataset. In both cases, I assume to have access to a line of densely sampled sources (as usually the case in ocean-bottom acquisition systems) and coarsely sampled receivers: both an irregular geometry (with receivers randomly jittered around a nominal coarse and regular geometry) and a coarse regular geometry are considered to test the interpolation capabilities of different measurement-guided reverse diffusion processes. More precisely, the available sources are sampled every 25m, whilst the receiver are originally sampled every 50m. The receiver array is further downsampled by a factor of 2, leading to a dataset with more or less 100m spacing between receivers (in the dithered scenario) and one with exactly 100m spacing between receivers (in the regular scenario). However, since the diffusion model has been trained on wavefields with 25m spacing, in both cases the output receiver grid is chosen to be sampled every 25m; in other words, I aim to interpolate the recorded data from 100m to 25m spacing.

Prior to training, the input data undergoes a series of processing steps to be amenable to the theory of diffusion models. First, the available common-receiver gathers are NMO corrected with a constant velocity of $2200m/s$ to mitigate aliasing effects in the steeply dipping events. Second, the amplitude weighting function previously described is estimated from and applied to each receiver gather. Next, patches of size $64 \times 64$ are extracted from each receiver gathers (the centers of the patches are selected over a grid with spacing of $32 \times 32$). However, since a very limited number of patches is likely to contain features corresponding to the strong direct arrival and refractions, the training data is augmented with a second sampling procedure that is carried out by selecting the patches' centers along the traveltime curve of the first arrival in each receiver gather. Finally, all patches are normalized between -1 an 1. Whilst the number of training samples varies from example to example, I always ensure to have at least 10.000 training samples in each of the following examples.

\begin{figure*}[!htb]
\centering
  \includegraphics[width=0.99\textwidth]{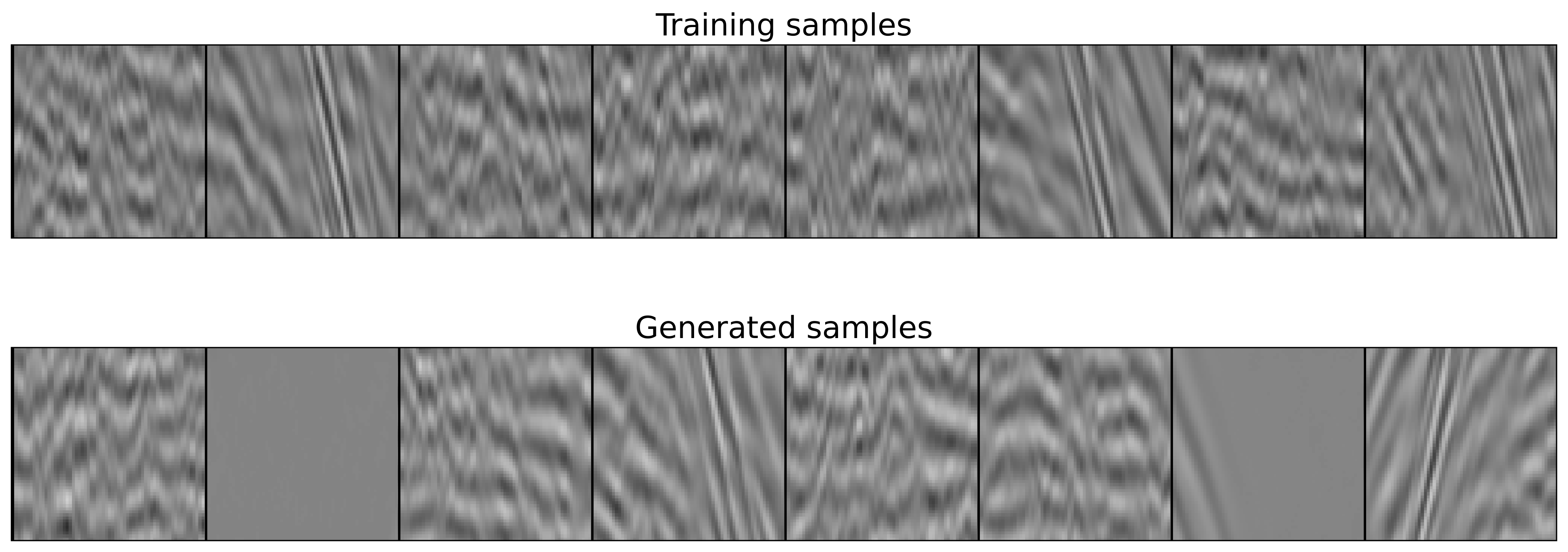}
  \caption{Top) Training samples, and bottom) unconditionally generated samples for the Volve synthetic training dataset.}
  \label{fig:uncgen_volvesynth}
\end{figure*}

The training process follows the standard procedure adopted when training diffusion models. A batch of patches is corrupted with random noise with variable standard deviation (as dictated by the chosen noise scheduler) and the network is tasked to predict the clean patches. In this example, two diffusion models are trained: the first uses a linear noise scheduler, whilst a cosine scheduler is used for the latter. After training, the denoising network can be embedded in both the unconditional and conditional sampling processes described above. Starting from the unconditional case, eight noise realizations of the same size of the training patches are fed into the network trained on the synthetic data, and after 1000 steps, realistic looking wavefields with features that resemble both linear and curved arrivals are produced (Figure \ref{fig:uncgen_volvesynth}). Whilst I show realizations generated from the diffusion model that uses the linear noise scheduler, results of similar quality can be obtained also from the diffusion model that uses the cosine noise scheduler.\\\\

\begin{figure*}[!htb]
\centering
  \includegraphics[width=0.99\textwidth]{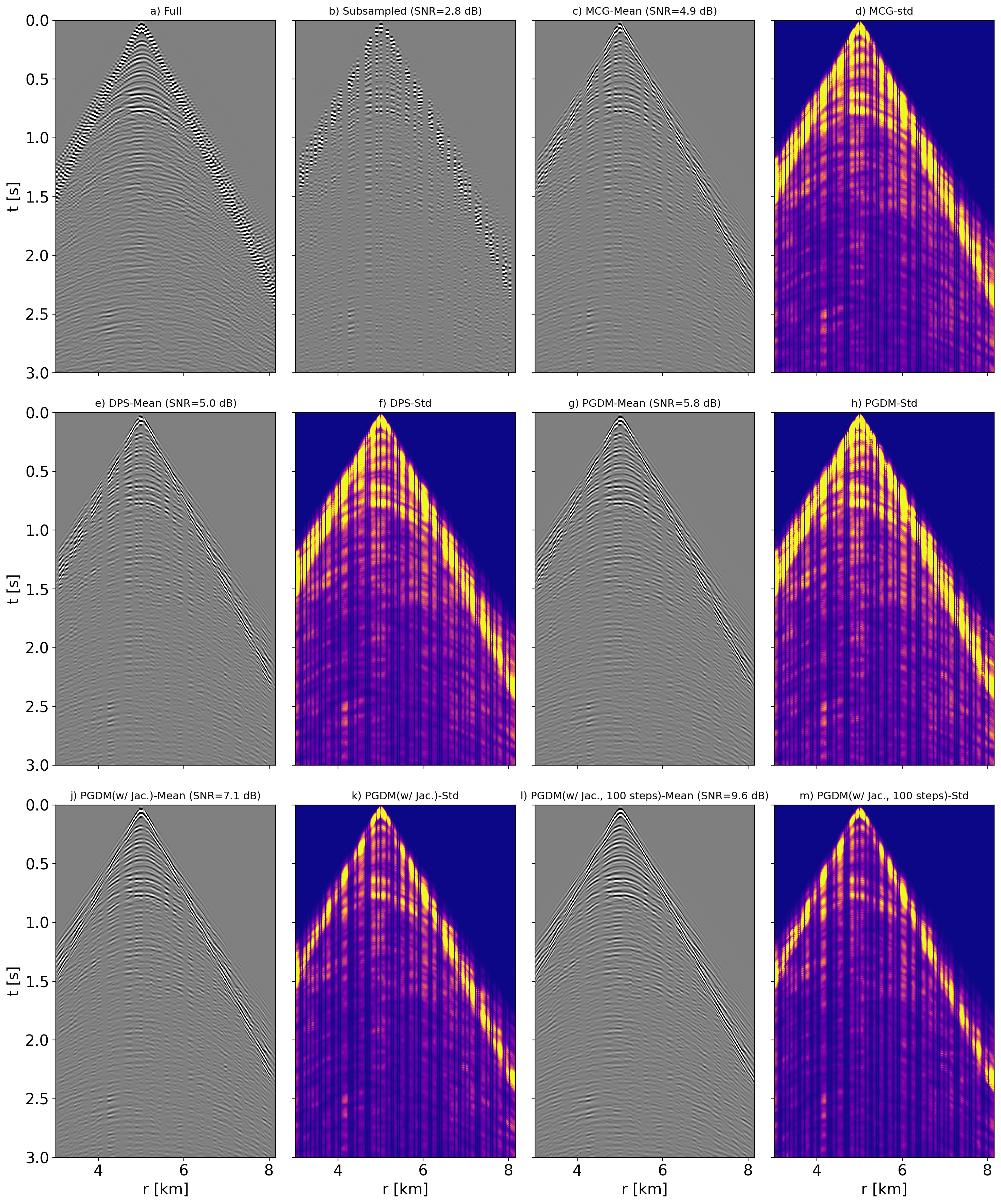}
  \caption{Mean and standard deviation of the reconstructed wavefield for different reverse diffusion processes in the case of jittered subsampling. a) True data (sampled every 50m), b) Subsampled data, c-d) MCG, e-f) DPS, g-h) PGDM with Jacobian, j-k) PGDM without Jacobian, l-m) PGDM without Jacobian and gapped reverse process.}
  \label{fig:int_volvesyth_dither_comparison}
\end{figure*}

\textbf{Volve synthetic - jittered subsampling}
After having verified the impressive generative capabilities of the  diffusion model, I aim now to use it as a prior in a seismic interpolation inverse problem. In order to do so, 50 random noise realizations of the size of an entire shot gather are created and used as starting guess for the measurement-guided sampling process. Figure \ref{fig:int_volvesyth_dither_comparison} shows the mean and standard deviation of a shot gather near the middle of the acquisition line for a series of experiments that I will discuss in the following, alongside the reference 'fully' sampled shot gather (i.e., sampled every 50m) and the subsampled shot gather that is used to guide the reverse diffusion process. To begin with, I consider the MCG and DPS samplers (Figures \ref{fig:int_volvesyth_dither_comparison}c-d and e-f). For noise-free inverse problems, like the seismic interpolation problem, the additional projection step into the measurement subspace performed by the MCG sampler is supposed to improve the converge behavior of the reverse process. This contribution is however shown to be not significant in my numerical results; MCG is actually performing poorer than DPS in terms of SNR ($\sim 0.1 dB$). The standard deviation is instead very similar for both cases, with small values (which should approach zero) in the available traces and large values in the missing traces. Next, I consider the PGDM sampler: in this case, since the interpolation operator is effectively a restriction operator, $\mathbf{G} \mathbf{G}^T=\mathbf{I}$. Moreover, since I assume to have noise-free data ($\sigma_w^2=0$), the pseudo-inverse term vanishes. Nevertheless, as shown in the numerical results, the scaling factor $\eta_t$ used in PGDM seems to be more effective than the one used by the DPS sampler (Figures \ref{fig:int_volvesyth_dither_comparison}g-h); as such, the overall SNR of this sampler is higher than that of the other two methods ($\sim 0.8 dB$). 

\begin{figure*}[!htb]
\centering
  \includegraphics[width=0.99\textwidth]
  {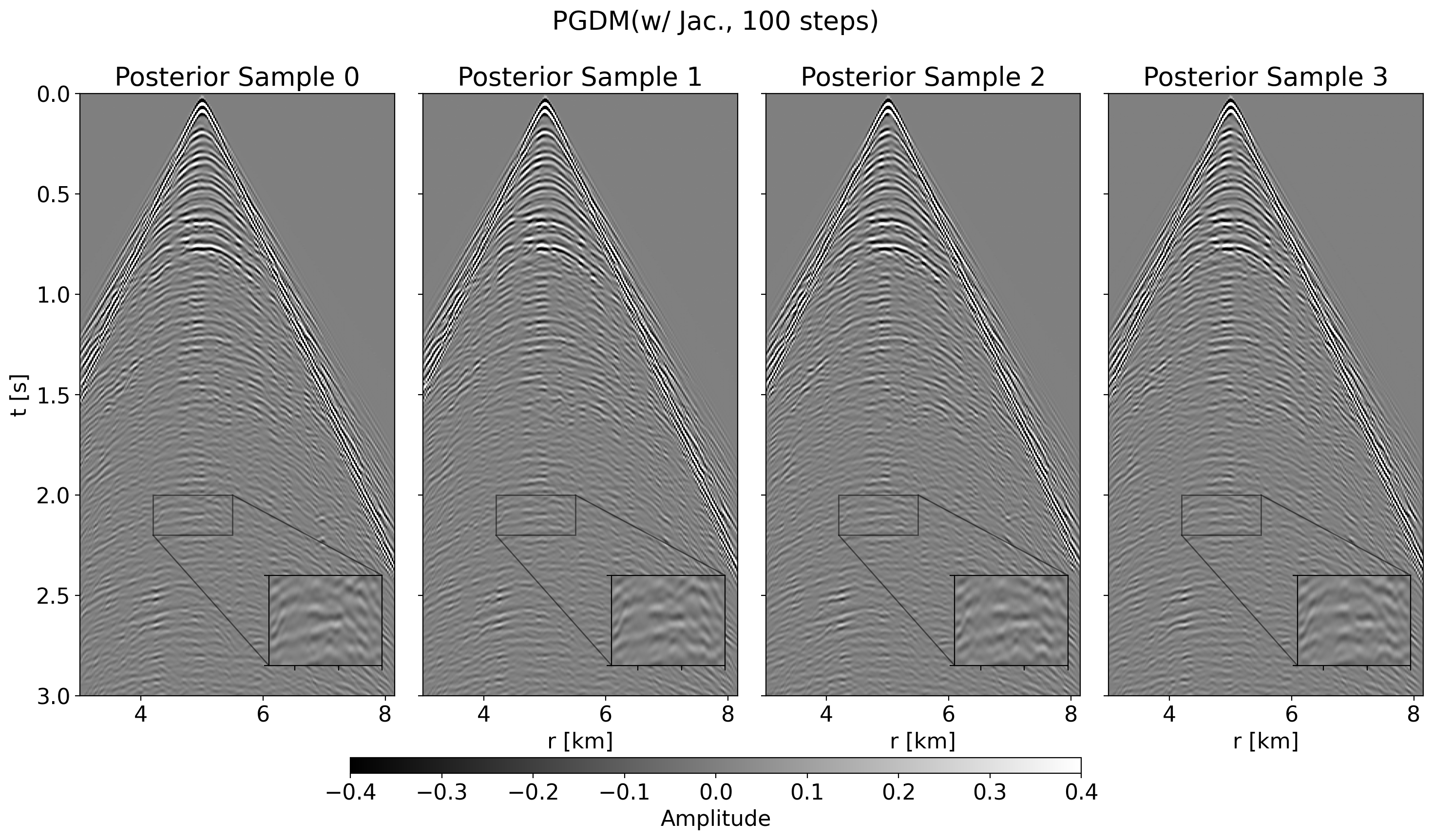}
  \caption{Four realizations of the wavefield reconstructed from irregularly sampled data using 100 steps of the PGDM sampler (without Jacobian and with a jump of 10 between each step).}
  \label{fig:int_volvesyth_dither_pdgmreal}
\end{figure*}

Next, I perform an experiment where I omit the contribution of the Jacobian of the denoiser (i.e., $\nabla_{\mathbf{x}_t}\hat{\mathbf{x}}_0) \approx \mathbf{I}$). Whilst this test was initially meant to reduce the computational cost of each reverse step, the performance of the overall reverse diffusion process seems to be boosted (from 5.8dB to 7.1dB as shown in Figure~\ref{fig:int_volvesyth_dither_comparison}j), suggesting that for this specific inverse problem the contribution of the Jacobian of the network does not seem to benefit the reconstruction process. Moreover, the standard deviation of the available traces is further reduced (Figure~\ref{fig:int_volvesyth_dither_comparison}k), sign of the fact that the reverse process has produced more representative samples of the posterior distribution. 

In all of the experiments up until now, the reverse process is warm-started with a noisy version of the available traces at step $T_s=200$, such that only 200 steps of denoising are required to obtain the final solution. This is a commonly adopted strategy in the literature to reduce the cost of the reverse diffusion process. Moreover, the best performing diffusion model is the one using a linear noise scheduler. However, as suggested in~\cite{Song2023}, for problems where a starting guess is not strictly required (as is the case in seismic interpolation), one can alternatively start the reverse diffusion process from the last step (i.e., $T_s=T=1000$) and skip some of the diffusion steps following the update rule in~\cite{kawar2022denoising}. In the last experiment, I apply this strategy performing a step every 10, meaning that the reverse diffusion process is initialized with purely random noise and only 100 steps are carried out. For simplicity, I refer to this procedure as \textit{gapped} reverse diffusion process from here onwards. The reconstructed wavefields are of much higher quality compared to those produced by any of the previous approaches, with the mean solution reaching an SNR of 9.1dB (Figure~\ref{fig:int_volvesyth_dither_comparison}l-m). From a visual standpoint the reconstructed events are more continuous with a much less pronounced decay in amplitude  within the reconstructed traces. This is also clearly visible in the individual realizations (Figure~\ref{fig:int_volvesyth_dither_pdgmreal}), and not only in the mean wavefield. 

\begin{figure*}[!htb]
\centering
  \includegraphics[width=0.8\textwidth]{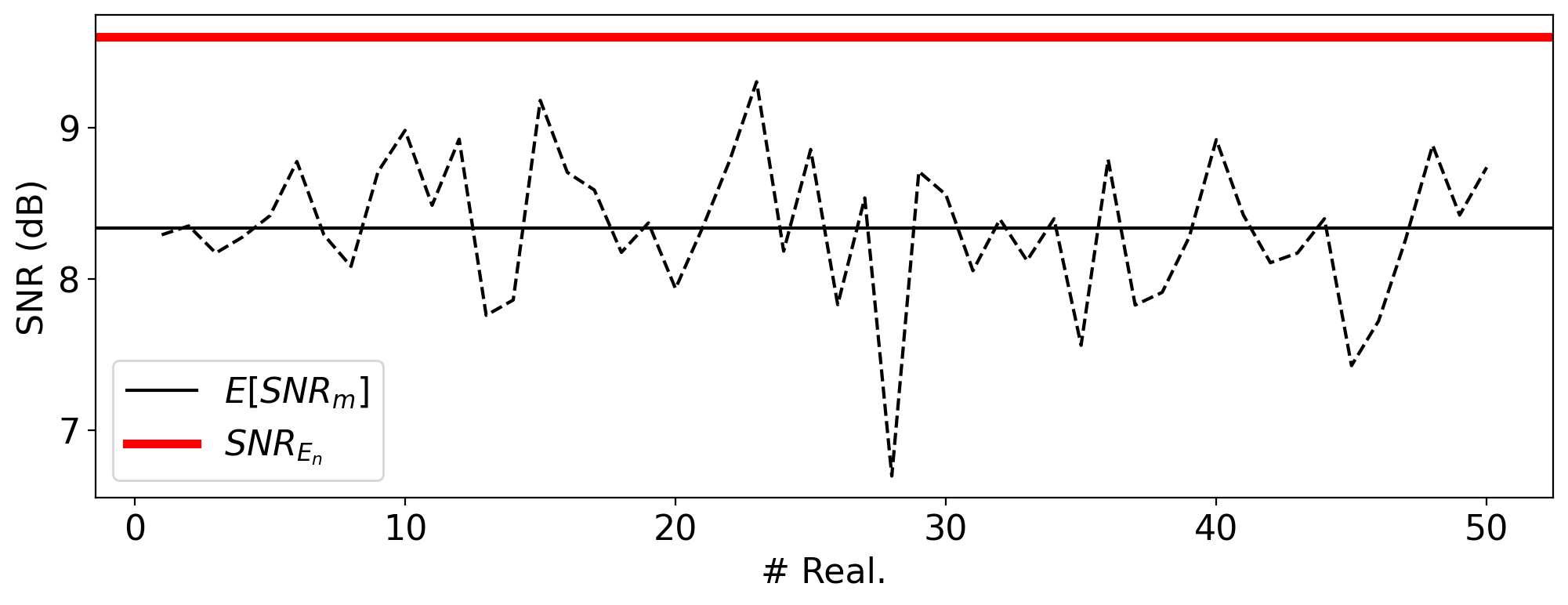}
  \caption{Signal-to-noise ratio comparison between each individual realization (dashed black line), average of the SNRs of each realization (black line), SNR of the wavefield obtained by averaging all of the individual realizations (red line).}
  \label{fig:int_volvesyth_dither_pdgmsnr}
\end{figure*}

Finally, in Figure~\ref{fig:int_volvesyth_dither_pdgmsnr} I show what I consider an interesting and perhaps (at first) puzzling result: given a number of samples obtained by repeatedly running the reverse diffusion process with a different random noise initialization, the overall SNR (obtained as the average of the SNR of each individual sample) is shown to be lower than the SNR of the mean solution (i.e., obtained by averaging the realizations). Moreover, the SNR of the mean solution is higher than that of most of the samples. This may seem odd at first: as we can see from Figures~\ref{fig:int_volvesyth_dither_comparison} and ~\ref{fig:int_volvesyth_dither_pdgmreal}, the amount of details shown in each realization is in fact somehow smeared in the average solution, leading to what our eyes perceive as a poorer result. However, as explained in~\cite{Blau2018}, this phenomenon is due to the so-called "perception-distortion trade-off", which arises in the context of denoising when the denoiser is trained by minimizing the MSE between the true and denoised result (i.e., MMSE denoiser). By doing so, the denoiser achieves optimal L2 distortion; however its perceptual quality is usually compromised. Since the measurement-guided reverse processes described in this work aim to solve inverse problems while striving for high perceptual quality, each individual realization is likely to be of lower SNR than their average, even though our visual perception of its quality is likely to be higher than that of the mean solution.\\

\begin{figure*}[!htb]
\centering
  \includegraphics[width=0.99\textwidth]{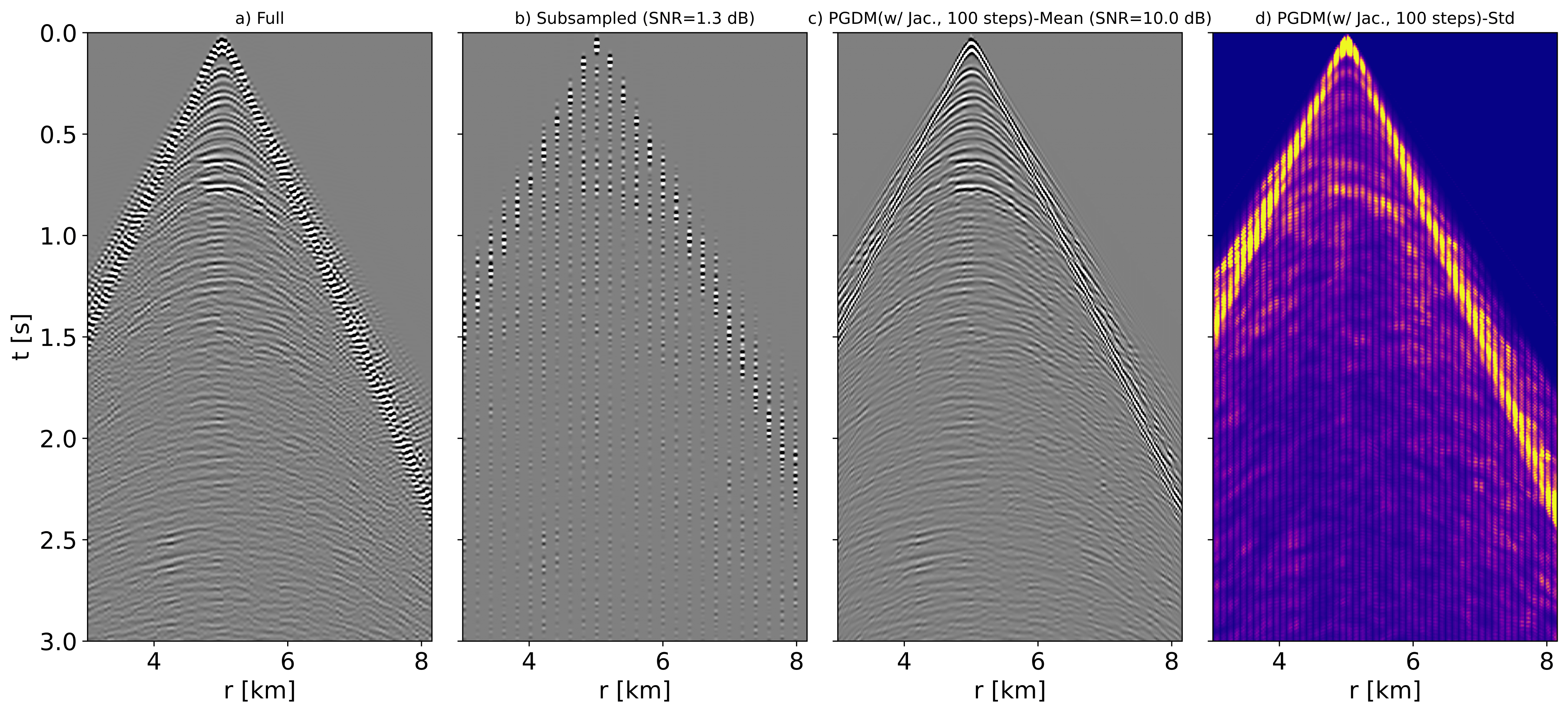}
  \caption{a) True data (sampled every 50m), b) Subsampled data (regular subsampling with a distance of 100m between traced), c-d) mean and standard deviation of the reconstructed wavefield (sampled every 25m) for the PGDM sampler.}
  \label{fig:int_volvesyth_reg_comparison}
\end{figure*}

\begin{figure*}[!htb]
\centering
  \includegraphics[width=0.99\textwidth]{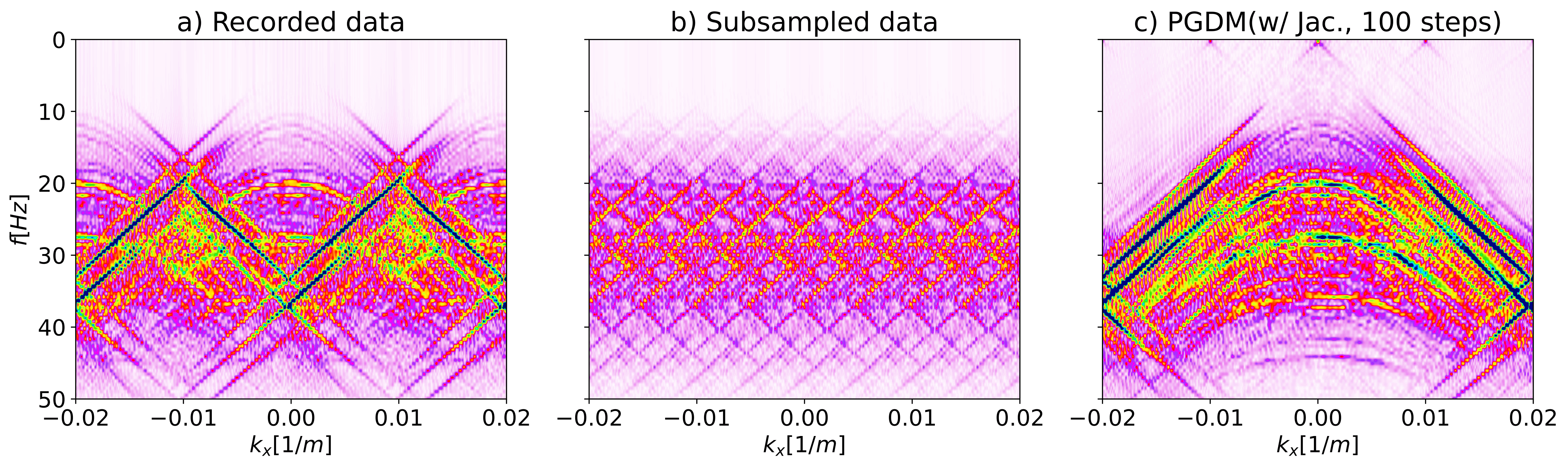}
  \caption{Fourier-wavenumber spectra for the a) recorded data, b) subsampled data, c) mean of the wavefields reconstructed using the PGDM sampler.}
  \label{fig:int_volvesyth_regular_fk}
\end{figure*}

\begin{figure*}[!htb]
\centering
  \includegraphics[width=0.99\textwidth]{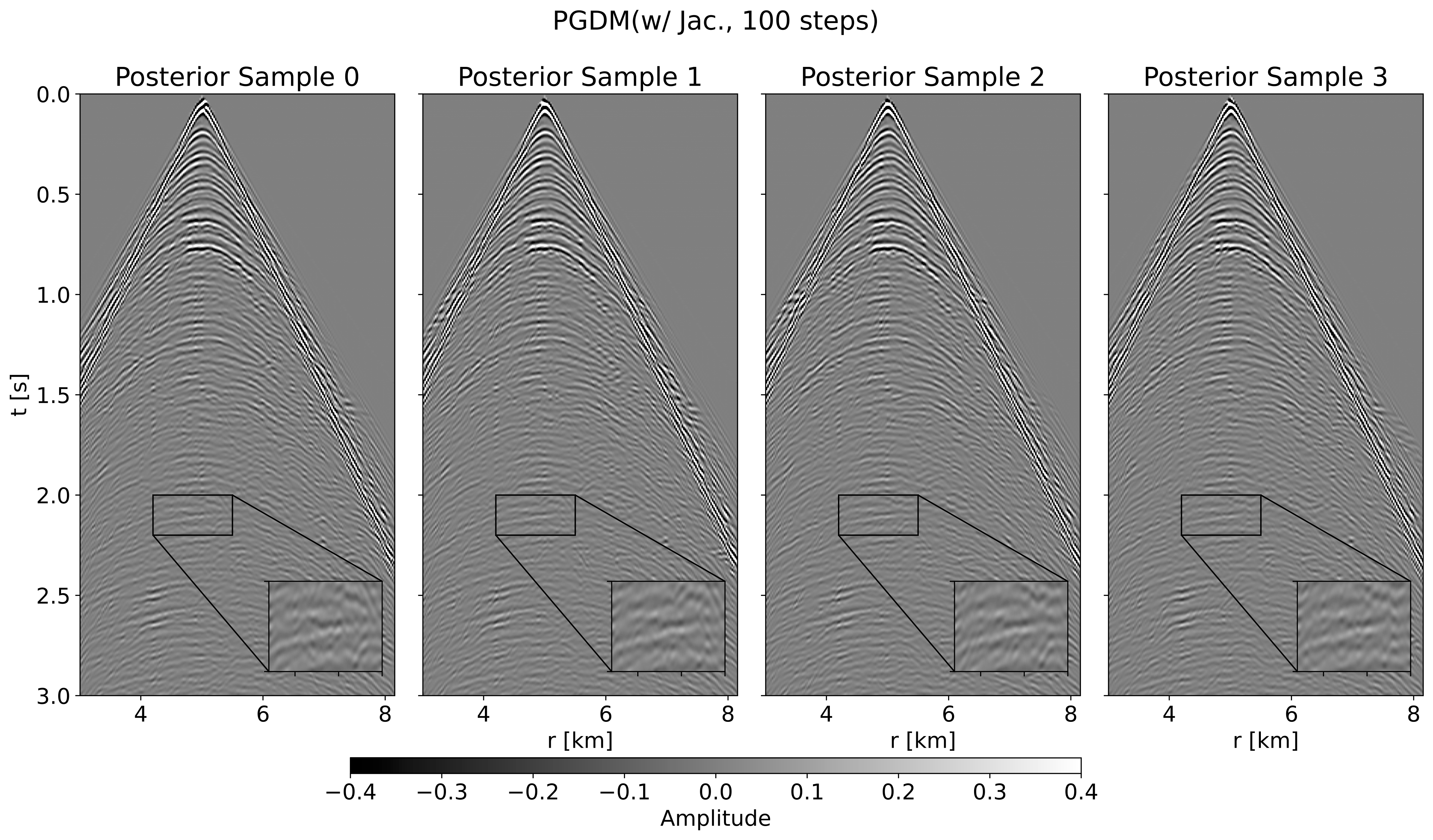}
  \caption{Four realizations of the reconstructed wavefield from regularly sampled data for the PGDM sampler (without Jacobian) and using 100 steps with a jump of 10 steps.}
  \label{fig:int_volvesyth_regular_pdgmreal}
\end{figure*}

\textbf{Volve synthetic - regular subsampling}
The same procedure is now repeated for the regularly subsampled data. In this case, I only consider the best performing strategy from the previous example, namely training the diffusion model with a cosine noise scheduler and applying the gapped PGDM sampler without Jacobian. The mean and standard deviation of the same shot gather used in the previous example, computed over 100 realizations, is displayed in Figure \ref{fig:int_volvesyth_reg_comparison}c-d. Once again, the reference 'fully' sampled shot gather (i.e., sampled every 50m) and the subsampled shot gather that is used to guide the reverse diffusion process are shown in panels a and b, respectively. Figure~\ref{fig:int_volvesyth_reg_comparison}c shows that underlying wavefield is successfully reconstructed despite the observed data is severely aliased (as also shown in Figure~\ref{fig:int_volvesyth_regular_fk}b). Finally, four realizations of the reconstructed waveforms are shown in Figure~\ref{fig:int_volvesyth_regular_pdgmreal}; note that whilst all of the reconstructed data are of high quality and show seismic-like features, some variability across realizations can be observed in both the main events at earlier times as well as in the coda of the wavefields.

\textbf{Volve field - jittered and regular subsampling}
To conclude this section, a series of similar experiments are conducted on a 2D line of the Volve OBC field dataset. To maintain consistency with the synthetic data, the input data is first low-passed with a cut-off frequency of 50Hz; note that this is however above the minimum frequency at which spatial aliasing manifests itself in the recorded data, and aliasing is further exacerbated by the additional subsampling process applied prior to interpolation. Moreover, only the sampling strategy that performed best in the synthetic case is adopted here. However, likely due to the increased complexity of the waveforms, 200 steps of the reverse diffusion process (with a jump of 5 steps) are required to achieve the results presented in the following. Moreover, an additional preconditioner is applied to the gradient, which removes any contribution outside of the signal cone in the frequency-wavenumber domain.

\begin{figure*}[!htb]
\centering
  \includegraphics[width=0.99\textwidth]{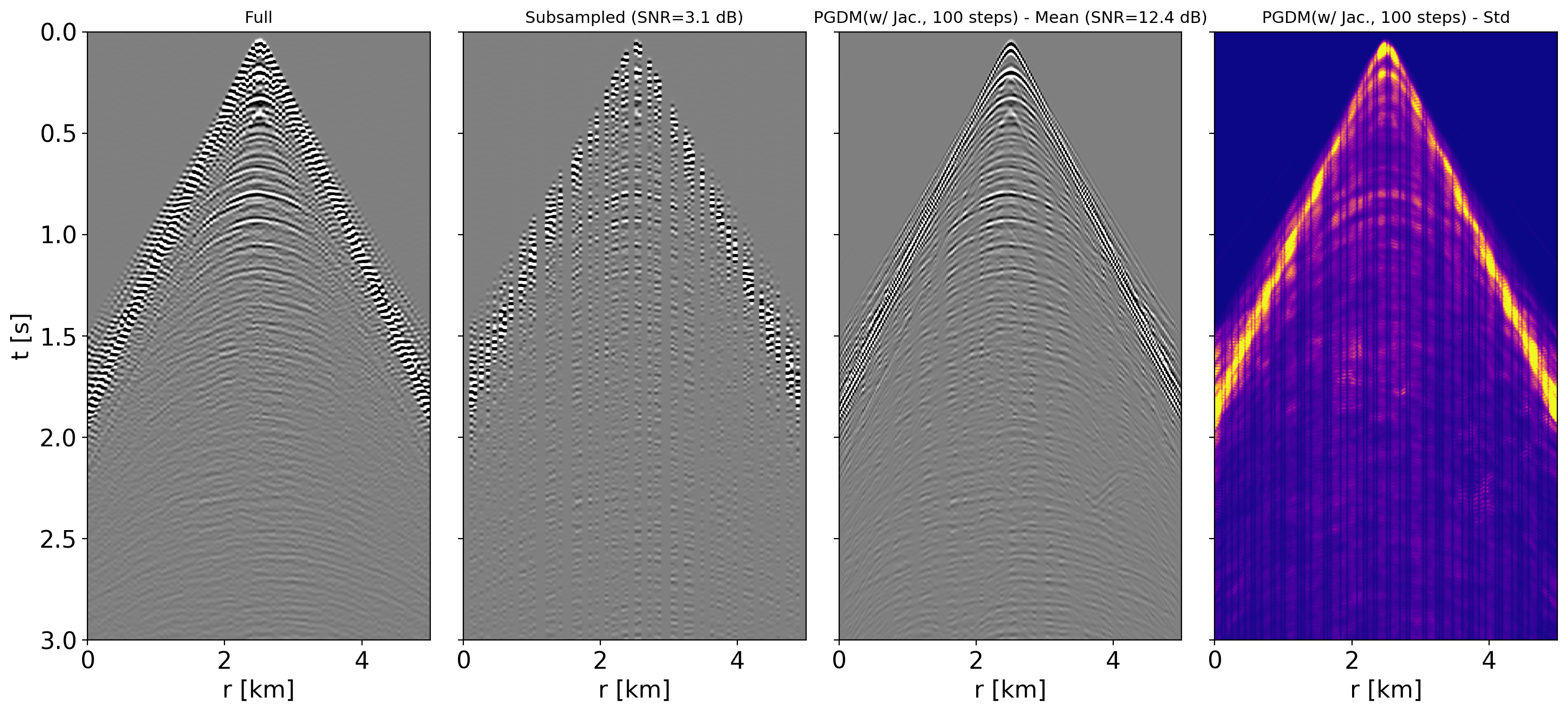}
  \caption{a) True data (sampled every 50m), b) Subsampled data, c-d) mean and standard deviation of the reconstructed wavefield (sampled every 25m) for the best performing sampler.}
  \label{fig:int_volve_dither_comparison}
\end{figure*}

\begin{figure*}[!htb]
\centering
  \includegraphics[width=0.99\textwidth]{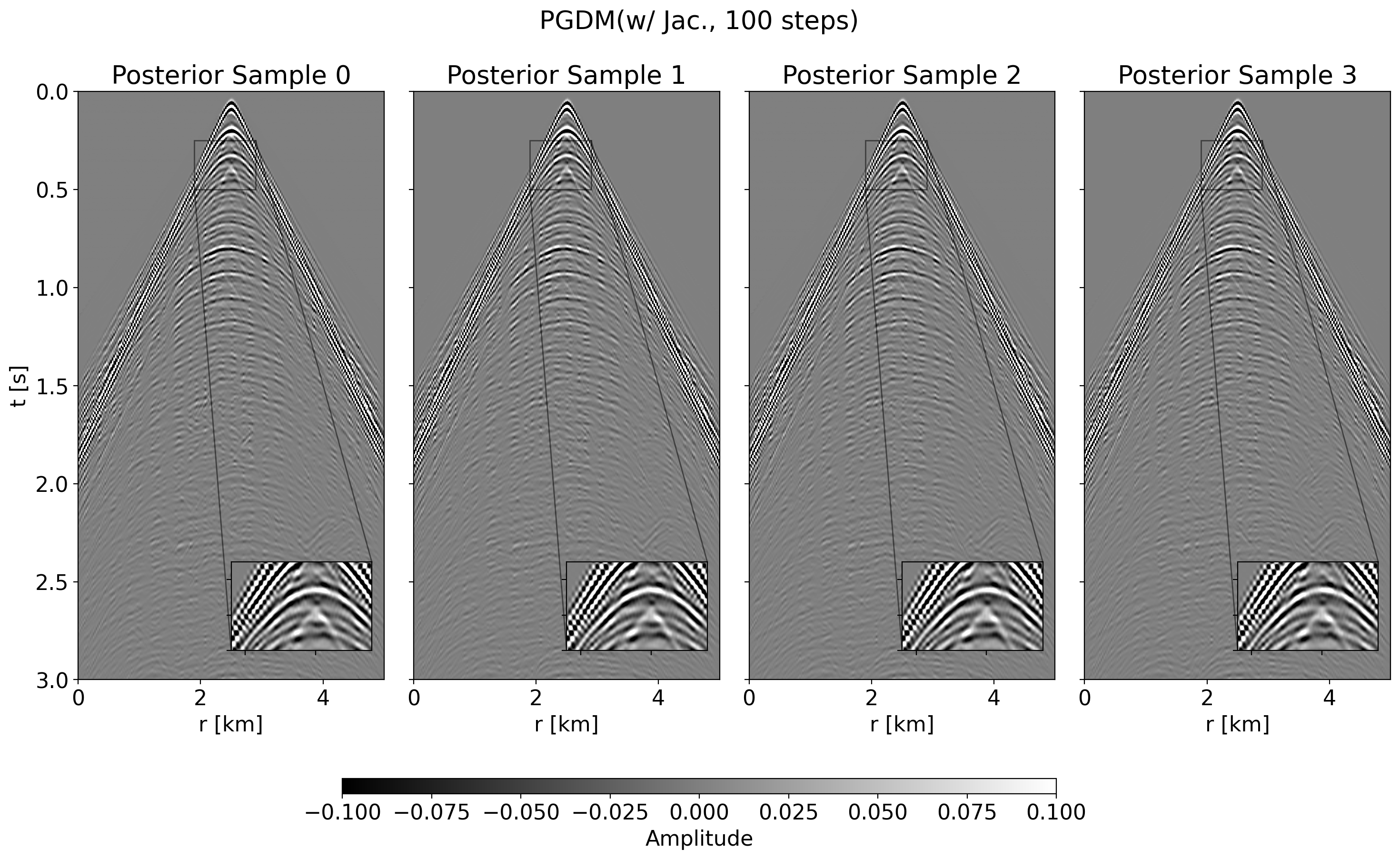}
  \caption{Four realizations of the reconstructed wavefield from irregularly sampled data for the gapped PGDM sampler.}
  \label{fig:int_volve_dither_pdgmreal}
\end{figure*}

\begin{figure*}[!htb]
\centering
  \includegraphics[width=0.99\textwidth]{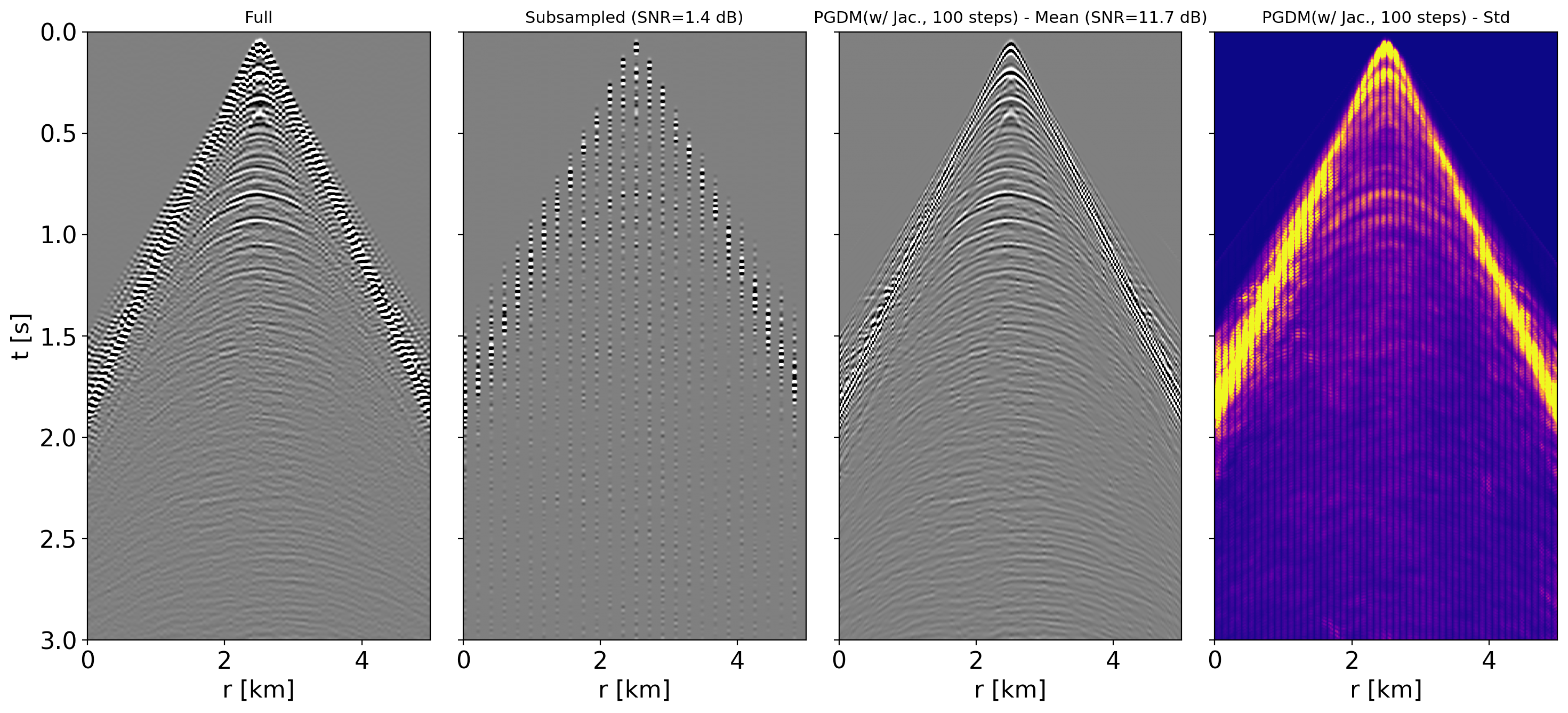}
  \caption{a) True data (sampled every 50m), b) Subsampled data (sampled every 100m), c-d) mean and standard deviation of the reconstructed wavefield (sampled every 25m) for the best performing sampler.}
  \label{fig:int_volve_regular_comparison}
\end{figure*}

\begin{figure*}[!htb]
\centering
  \includegraphics[width=0.99\textwidth]{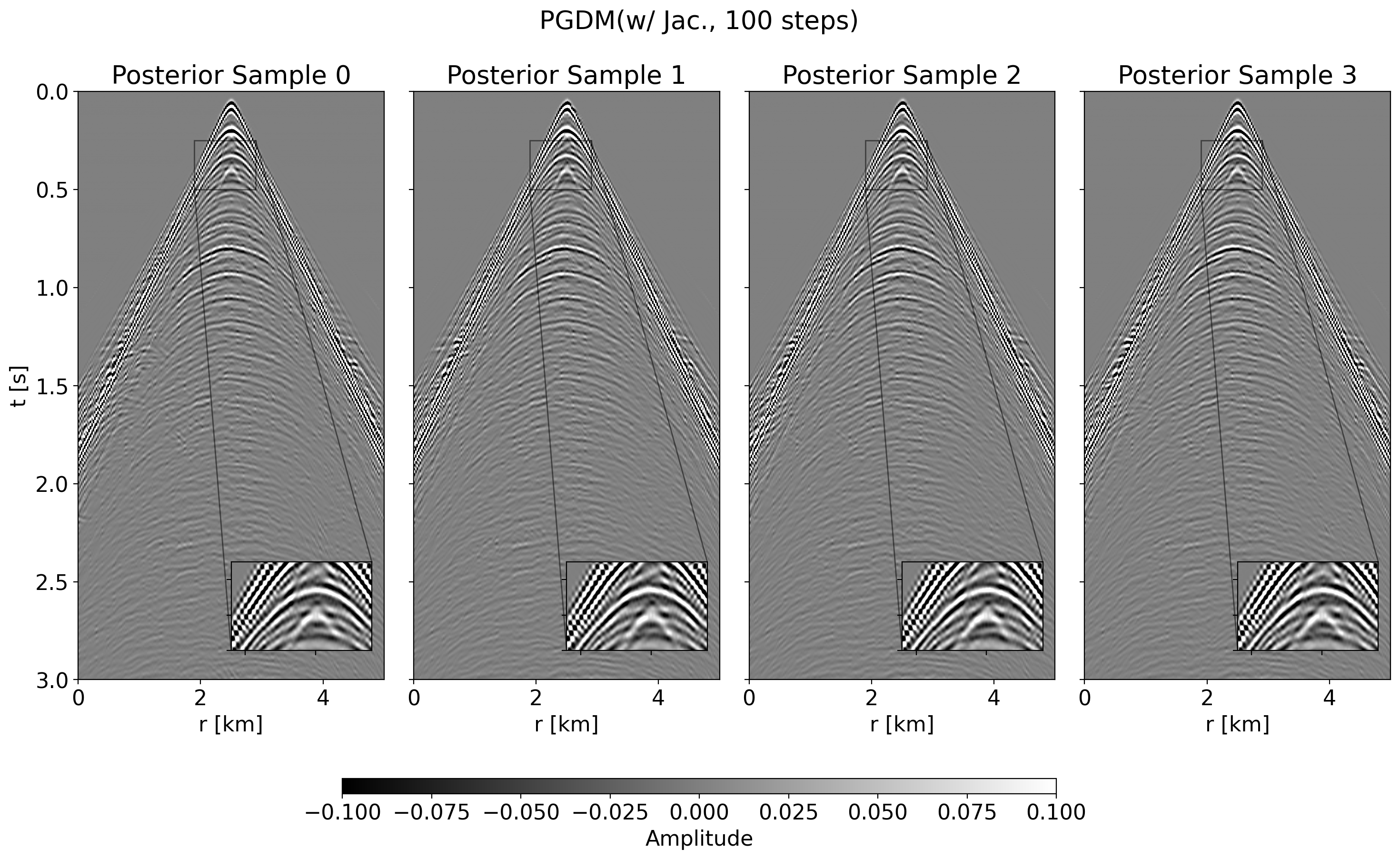}
  \caption{Four realizations of the reconstructed wavefield from regularly sampled data for the gapped PGDM sampler.}
  \label{fig:int_volve_regular_pdgmreal}
\end{figure*}

Starting from the jittered subsampling scenario, Figure~\ref{fig:int_volve_dither_comparison} shows the original data and the irregularly subsampled data from a source in the middle of the receiver line, alongside the mean and standard deviation of the reconstructed wavefield (computed over 100 realizations). Whilst the visual quality of the reconstruction is slightly poorer than the one of the synthetic data, with slight amplitude fluctuations along the spatial axis (i.e., smaller amplitudes appear in the reconstructed missing traces), the events with strongest amplitude and larger dips are still well reconstructed. Similarly, individual realizations show high degree of realism, yet present variability in the areas of missing traces, which are consistent with the stochastic nature of the reverse diffusion process Figure~\ref{fig:int_volve_dither_pdgmreal}. Similarly, when the data is regularly subsampled by a factor of two (leading to an overall distance between traces of 100m), the reverse diffusion process is capable of effectively reconstructing the missing traces despite severe spatial aliasing is observed in the frequency-wavenumber spectrum. Figures~\ref{fig:int_volve_regular_comparison} and ~\ref{fig:int_volve_regular_pdgmreal} show the mean and standard deviation of the reconstructed wavefield and four individual realizations, respectively.

\subsection{Post-stack Inversion}
In this section, I consider another popular linear inverse problem, namely post-stack seismic inversion. Compared to the interpolation scenario, the training dataset cannot be directly extracted from recorded data in this case. Instead, two different datasets are created using:
\begin{itemize}
    \item Four openly available geological models, namely the Marmousi model~\cite{Brougois1990}, the 1997 BP 2.5D migration benchmark model (also known as the Amoco model --~\cite{Etgen1998}, a 2D slice of the Seam Phase I  model~\cite{Fehler2011}, and a 2D slice of the EAGE/SEG Overthrust model~\cite{Brougois1990}. Note that in all cases, the acoustic impedance model needed by the post-stack modeling operator is created directly from the velocity model (assuming a constant density model). However, one could also use the density model (when provided) to create the acoustic impedance model.
    \item Five classes of the OpenFWI dataset~\cite{Deng2022}, namely the FlatVel-A, FlatFault-A, FlatFault-B, CurveFault-A, and Style-A classes. Once again, since only velocity models are provided, I assume here a constant density model to create the acoustic impedance model needed by the post-stack modeling operator.
\end{itemize}

Starting from these two datasets, six different subsets are generated and used to train six diffusion models. Random patches of size $64\times64$ are extracted from the openly available models, whilst the OpenFWI models, whose original size is $70\times70$, are cropped to also become of size $64\times64$. The six subsets are created as follows:
\begin{itemize}
    \item Marmousi: 1.000 patches are extracted from the Marmousi model;
    \item Marmousi+other: 2.000 patches are extracted from each of the models, for a total of 10.000 training samples;
    \item Other: 2000 patches are extracted from each of the models except the Marmousi model, for a total of 8.000 training samples;
    \item OpenFWI: 1200 patches are extracted from each of the five classes for a total of 6.000 training samples;
    \item OpenFWI-fault: 2000 patches are extracted from the FlatFault-A, FlatFault-B, and CurveFault-A, classes, for a total of 6.000 training samples;
    \item OpenFWI-nostyle: 1500 patches are extracted from the FlatVel-A, FlatFault-A, FlatFault-B, and CurveFault-A classes for a total of 6.000 training samples;
\end{itemize}
After training, the generative capabilities of the different diffusion models are evaluated by creating acoustic impedance models of size $64\times64$ as well as larger models of size $256\times128$ (where 256 refers to the samples along the lateral direction) -- see Appendix A. Note how different training datasets, although all consisting of synthetic (yet plausible) geological models, can strongly affect the prior information that is captured by the diffusion model -- and later injected into the inversion.\\

\begin{figure*}[!htb]
\centering
  \includegraphics[width=0.7\textwidth]{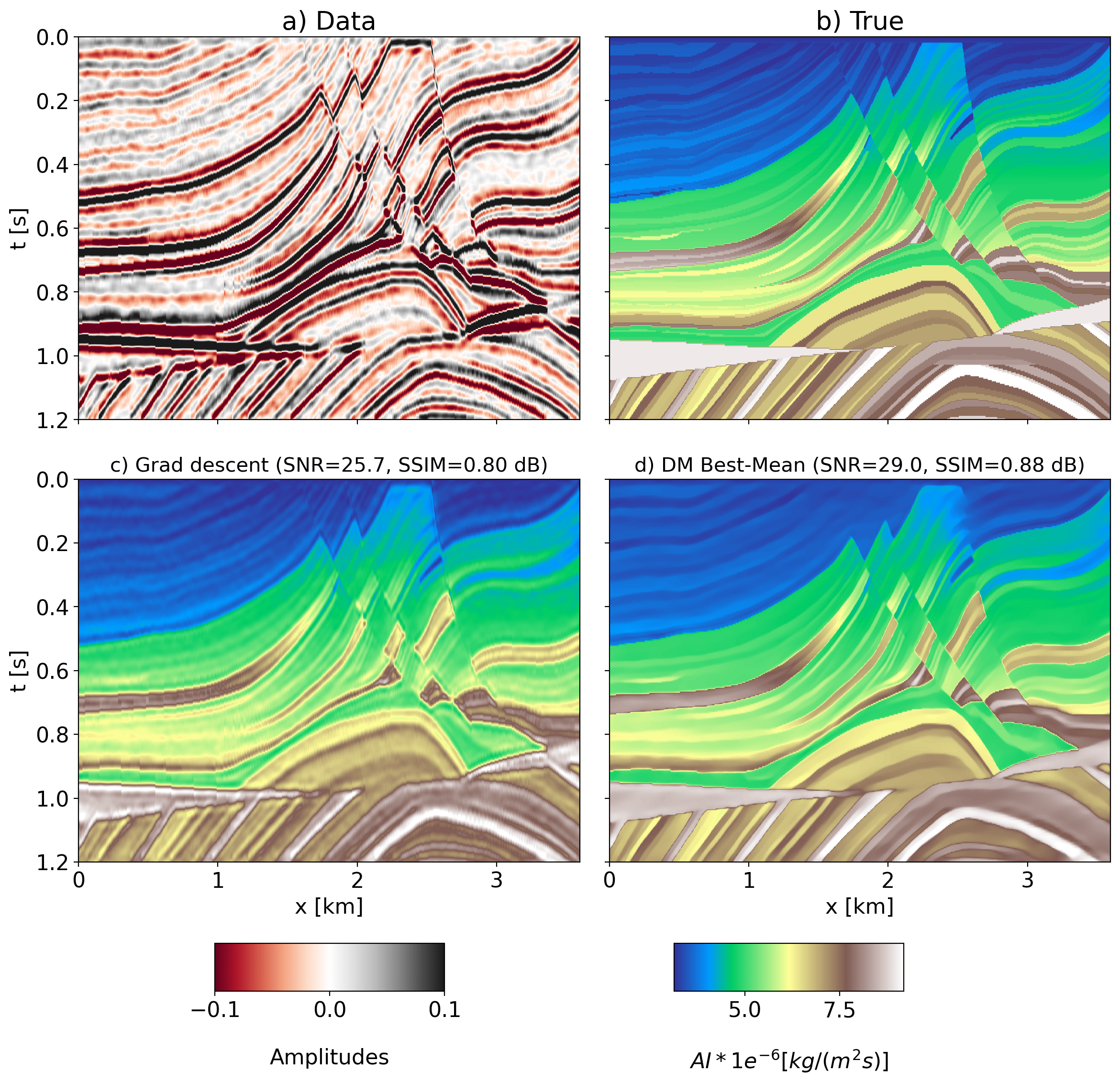}
  \caption{a) Data, b) True subsurface model, c) gradient descent benchmark solution, and d) mean of the best performing reverse diffusion process (i.e., PGDM).}
  \label{fig:post_marm_comparison}
\end{figure*}

\textbf{Marmousi}
In the first experiment, I model a post-stack seismic data from the Marmousi acoustic impedance model (Figure~\ref{fig:post_marm_comparison}b). The vertical axis is assumed here to represent two-way traveltime, and the convolutional modelling operator uses a Ricker wavelet with frequency equal to 15 Hz. Moreover, to mimic a realistic scenario, temporally and spatially band-passed noise is added to the data (Figure~\ref{fig:post_marm_comparison}a). This data is initially inverted using a vanilla gradient descent algorithm with fixed step-size for a total number of iterations equal to 200 (Figure~\ref{fig:post_marm_comparison}c); whilst a more accurate acoustic impedance model could be obtained using state-of-the-art iterative solvers with smoothness or Total Variation (TV) regularization, this choice is motivated by the fact that one could interpret the gradient descent algorithm as a bare-minimum version of any of the reverse diffusion processes used in this work, where only the last term of the update equation is used (i.e., $\mathbf{x}_s=\mathbf{x}_t - \eta_t \mathbf{g}_t$). In other words, this allows one to understand how the interaction between the score of the likelihood term (as used in the gradient descent algorithm) and the other terms used in the reverse diffusion process improves the final reconstructed model versus considering only the likelihood term. The mean of the best performing reverse diffusion process is finally shown in Figure~\ref{fig:post_marm_comparison}d. This is obtained by running the PGDM algorithm for 200 steps ($T_s = 200)$ starting from a heavily smoothed version of the true acoustic impedance model. One can immediately observe an improvement in SNR of about 3 dB for the measurement-guided reverse diffusion process compared to the gradient descent solution; this can be loosely quantified as the contribution of a carefully trained implicit prior. Note that this result represents an upper bound in terms of quality of reconstruction, as it relies on a diffusion model trained using patches extracted from the Marmousi acoustic impedance model. Whilst this does not represent a realistic scenario, in that one hardly has access to training data that are so close to the actual solution of the inverse problem at hand, it allows us to further investigate the role that different reverse diffusion processes and training datasets have in terms of the quality of the inversion process.

\begin{figure*}[!htb]
\centering
  \includegraphics[width=0.99\textwidth]{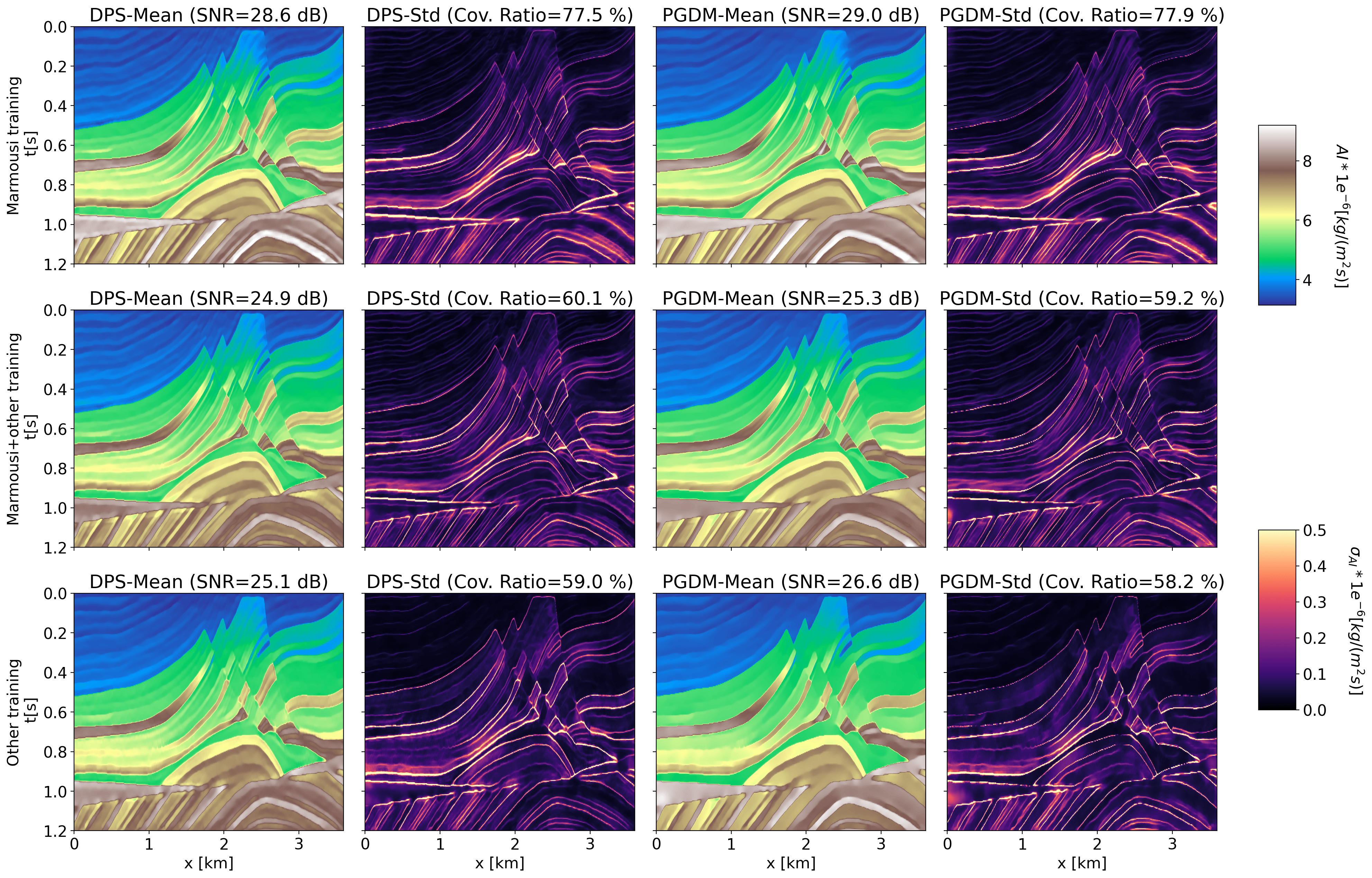}
  \caption{Mean and standard deviation for the DPS and PGDM diffusion sampling processes using training data based on the openly available velocity models.}
  \label{fig:post_marm_openmodels}
\end{figure*}

\begin{figure*}[!htb]
\centering
  \includegraphics[width=0.99\textwidth]{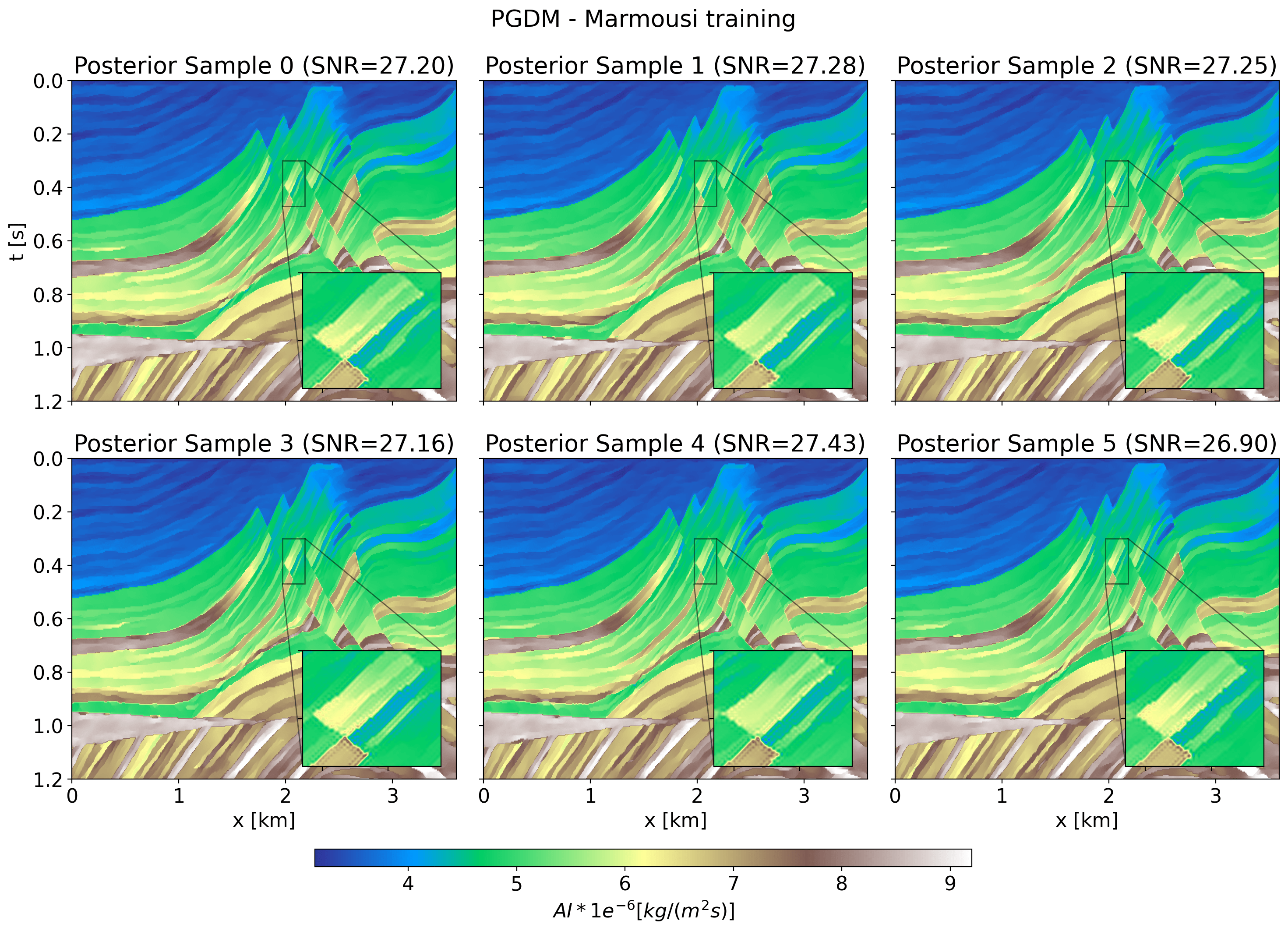}
  \caption{Realizations for the best performing measurement-guided diffusion sampling process.}
  \label{fig:post_marm_dpsreals}
\end{figure*}

Starting from the reverse diffusion process, the reconstruction capabilities of the DPS algorithm are now compared to those of the PGDM algorithm (top row of Figure~\ref{fig:post_marm_openmodels}); in both cases, the reverse diffusion process is started from $T_s = 200$ using a smoothed version of the true model. Moreover, 5 iterations of conjugate gradient are used to approximate the pseudo-inverse in the PGDM algorithm. Finally, compared to the seismic interpolation case, in the PGDM update for the seismic inversion problem, one must select the standard deviation of the noise $\sigma_w$; in my experiments, since the noise added to the data presents spatio-temporal correlation, $\sigma_w$ is selected to be half of the actual standard deviation of the noise. This is important to ensure that the diagonal term in the pseudo-inverse (i.e., $\sigma_w^2 \mathbf{I}$) does not dominate the other term (i.e., $r_t^2\textbf{G}\textbf{G}^T$), otherwise the PGDM solutions tends to approach that of DPS.

Whilst both algorithms produce a satisfactory solution and a similar estimate of uncertainty, the PGDM algorithm seems to produce a mean model that is slightly better in terms of resolution and overall amplitude fidelity for all the training datasets, likely due to the contribute of the pseudo-inverse operator. For both algorithms, on the other hand, the resulting standard deviation reveals that the main uncertainties in the inverse process are related to the boundaries (or interfaces) between layers. This is a reasonable outcome as the trained diffusion model implicitly constraints the solution to be piece-wise continuous in agreement with the training data; as such, the main differences between different realizations are concentrated near the boundaries. Finally, by looking at four different realizations of the PGDM algorithm (Figure~\ref{fig:post_marm_dpsreals}), it is evident how such an algorithm is able to produce highly realistic samples from the posterior distribution with small geologically plausible variations, as further highlighted in the close-ups. This is a major feature of the diffusion models described in this work, which totally differ from other state-of-the-art Bayesian approaches to inverse problems that struggle to produce realistic looking realizations (despite estimating accurate statistical moments - e.g., mean and standard deviation). Examples of methods that belong to this class are Monte Carlo Markov Chain (MCMC) algorithms~\cite{James2012, Gebraad2020, Berti2024}, mean-field Variational Inference methods (e.g., ~\cite{ZHANG202173}), the Stein Variational Gradient Descent (SVGD) method~\cite{Zhang2020, Izzatullah2024}, as well as recently developed uncertainty quantification techniques with Implicit Neural Representations~\cite{Romero2024}.

\begin{figure*}[!htb]
\centering
  \includegraphics[width=0.99\textwidth]{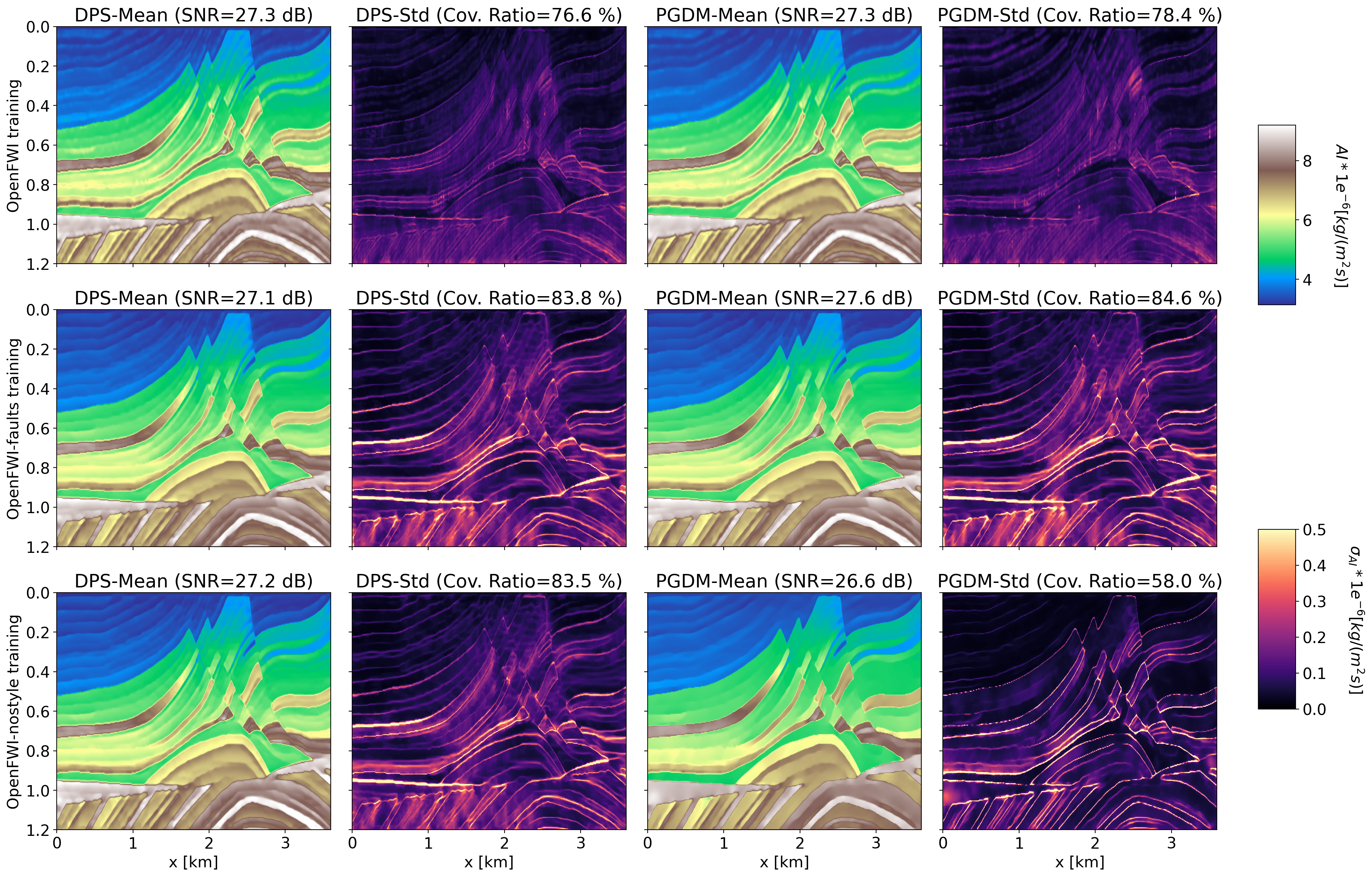}
  \caption{Mean and standard deviation for the DPS and PGDM diffusion sampling processes using training data based on the openFWI velocity models.}
  \label{fig:post_marm_openfwi}
\end{figure*}

\begin{figure*}[!htb]
\centering
  \includegraphics[width=0.99\textwidth]{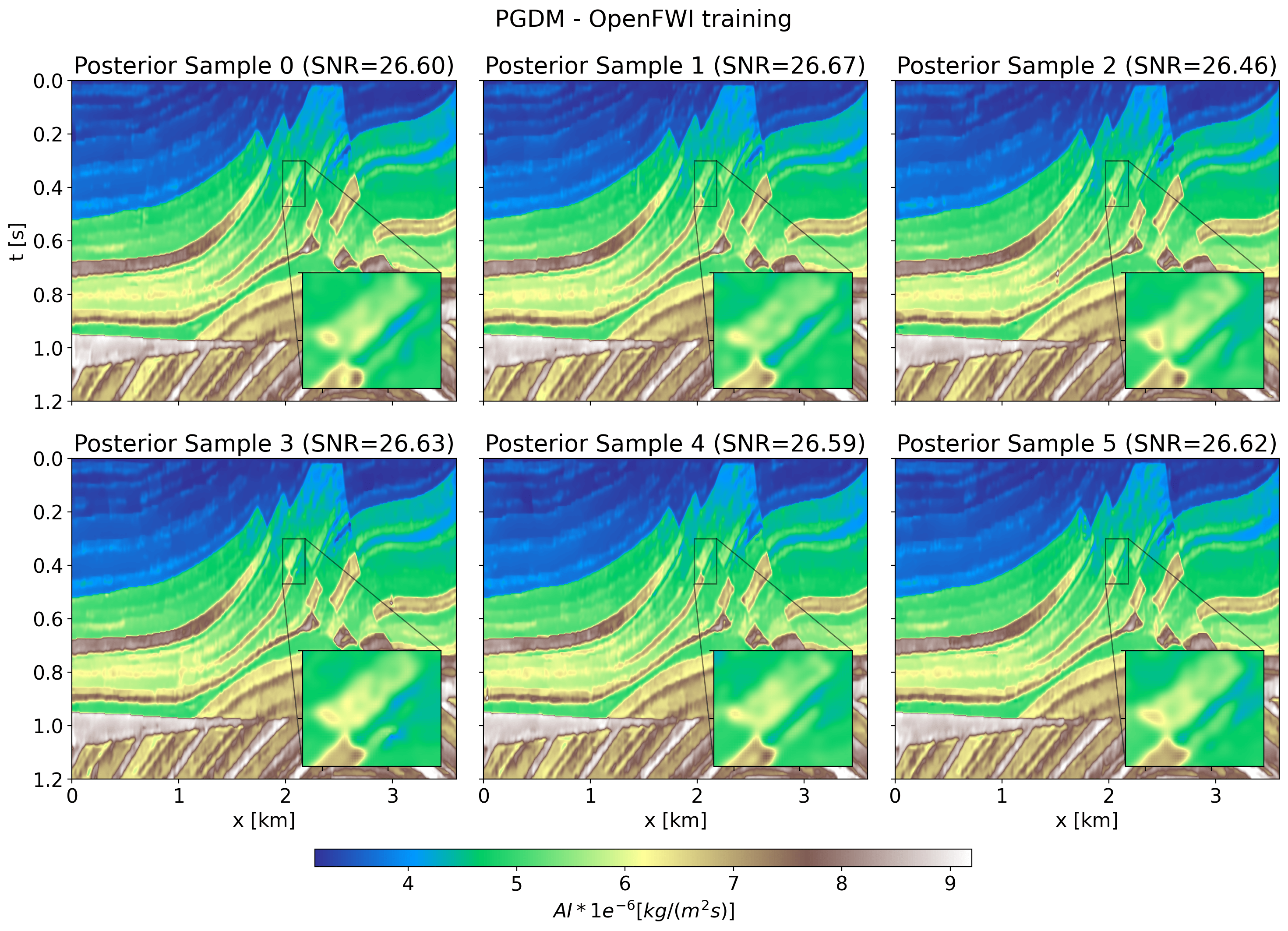}
  \caption{Realizations obtained using the diffusion model trained on the OpenFWI dataset.}
  \label{fig:post_marm_pgdmreals_openfwi}
\end{figure*}

\begin{figure*}[!htb]
\centering
  \includegraphics[width=0.99\textwidth]{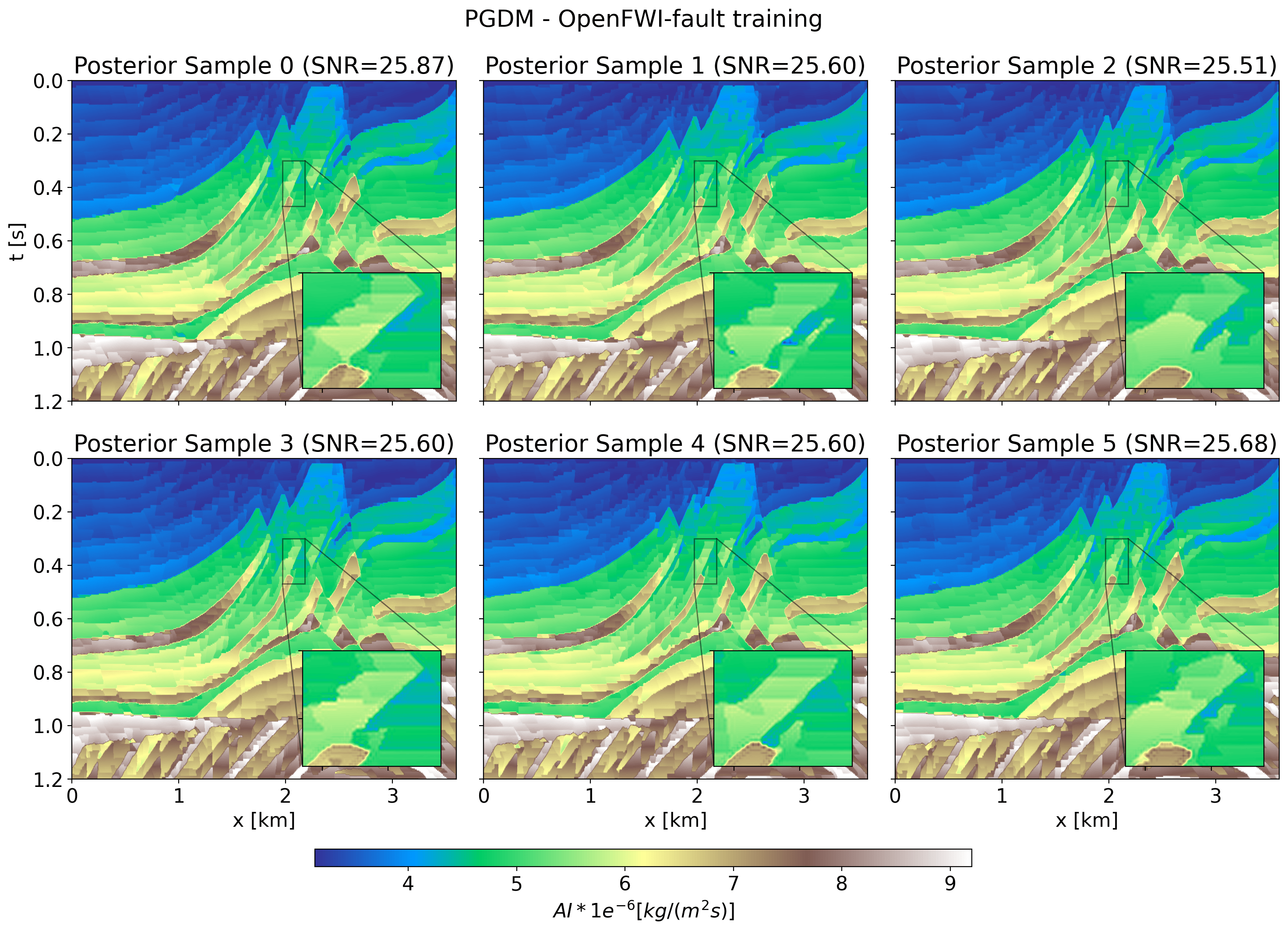}
  \caption{Realizations obtained using the diffusion model trained on the OpenFWI-fault dataset.}
  \label{fig:post_marm_pgdmreals_openfwi1}
\end{figure*}

\begin{figure*}[!htb]
\centering
  \includegraphics[width=0.7\textwidth]{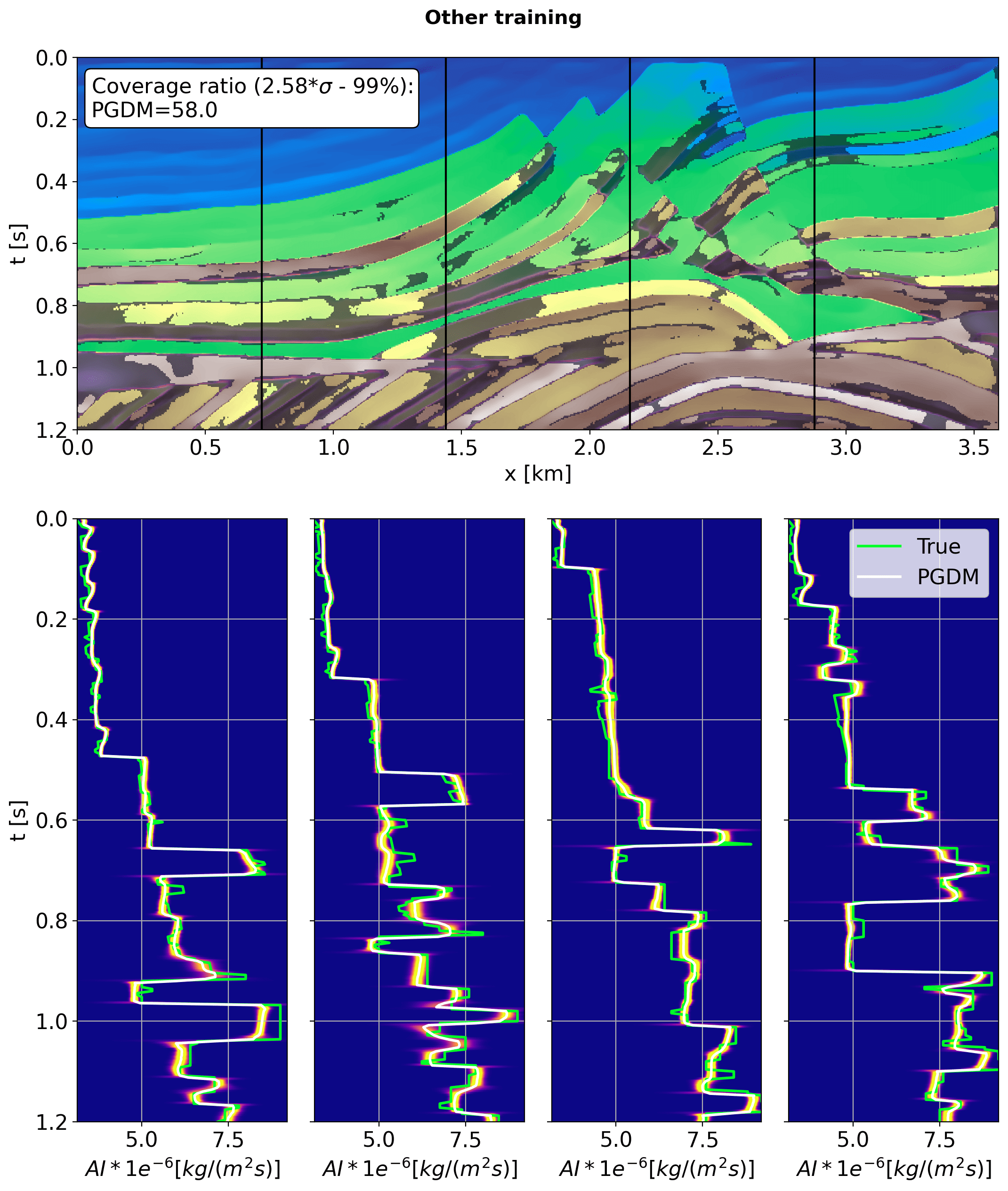}
  \caption{Marginal distributions along 4 vertical profiles for the PGDM sampler using the diffusion model trained with the Other training dataset. In the top panel, the mean model is co-rendered with the associated standard deviation.}
  \label{fig:post_marm_openmodels_1dstat}
\end{figure*}

\begin{figure*}[!htb]
\centering
  \includegraphics[width=0.7\textwidth]{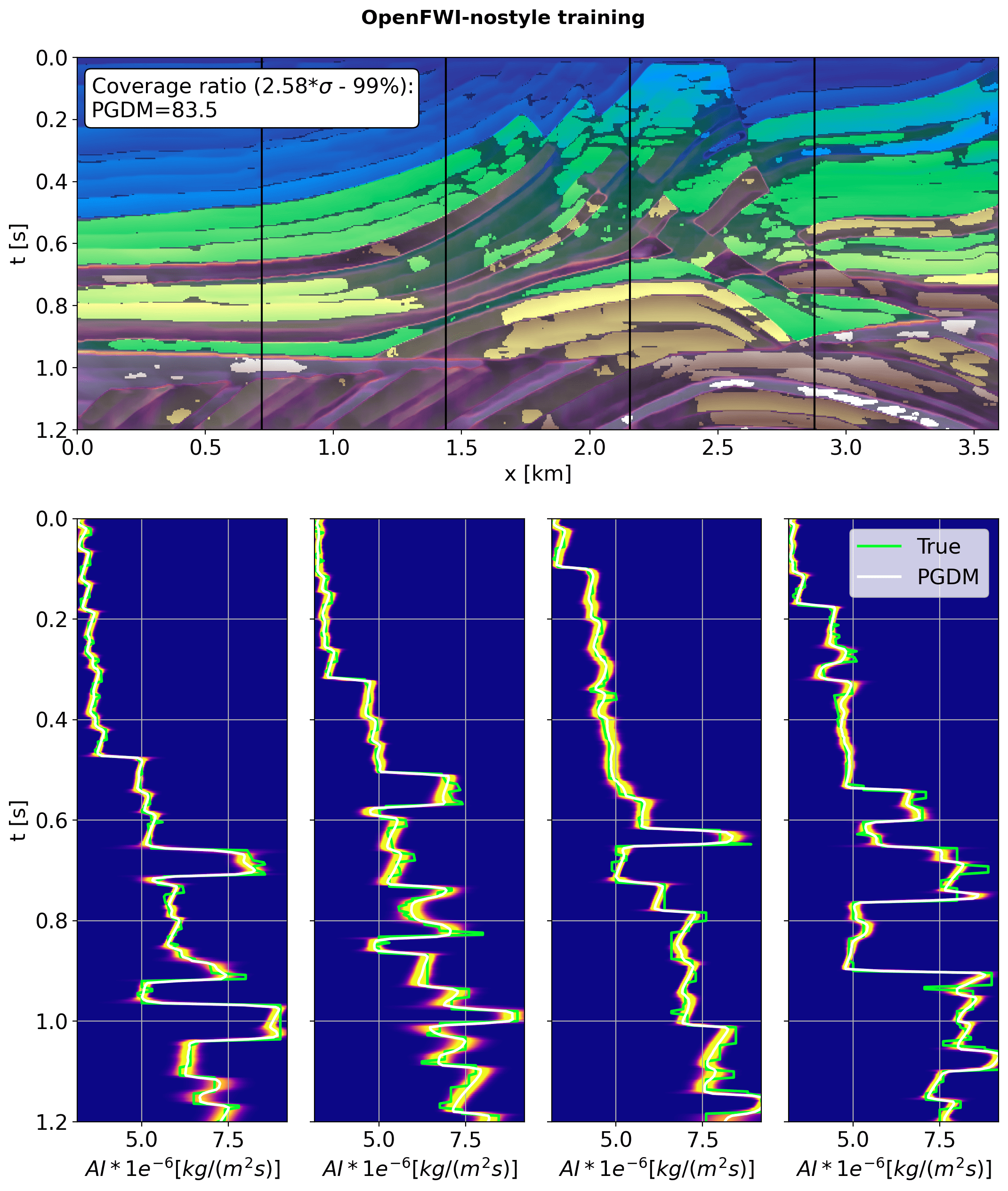}
  \caption{Marginal distributions along 4 vertical profiles for the PGDM sampler using the diffusion model trained with the OpenFWI-nostyle training dataset. In the top panel, the mean model is co-rendered with the associated standard deviation.}
  \label{fig:post_marm_openfwi_1dstat}
\end{figure*}

\begin{figure*}[!htb]
\centering
  \includegraphics[width=0.7\textwidth]{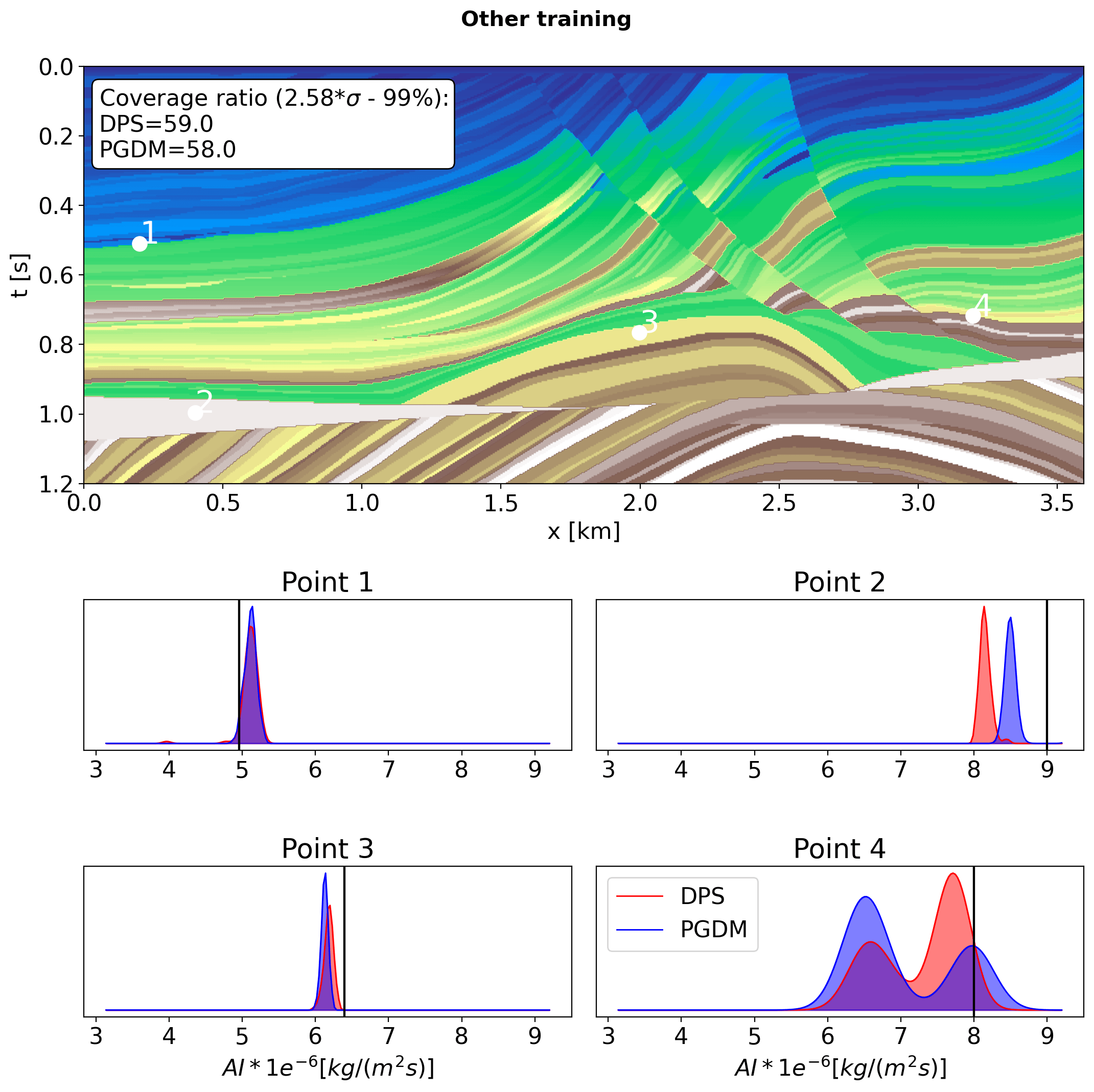}
  \caption{Marginal distributions at 4 different locations in the subsurface (displayed on top of the true model) for the DPS and PGDM samplers using the diffusion model trained with the Other training dataset. Black vertical lines represent the true values at the four locations.}
  \label{fig:post_marm_openmodels_marginals}
\end{figure*}

\begin{figure*}[!htb]
\centering
  \includegraphics[width=0.7\textwidth]{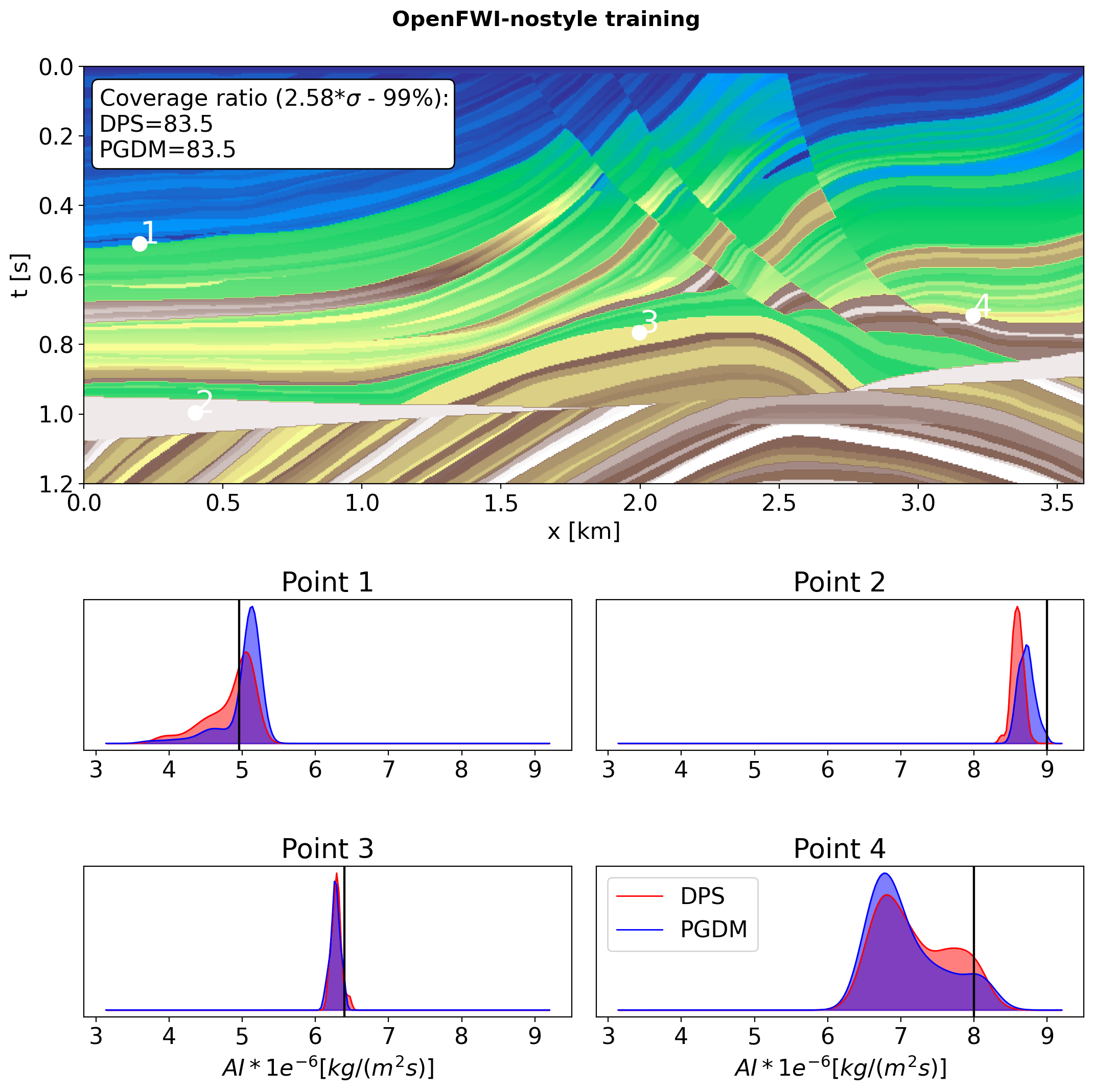}
  \caption{Marginal distributions at 4 different locations in the subsurface for the DPS and PGDM samplers using the diffusion model trained with the OpenFWI-nostyle training dataset. Keys as in Figure~\ref{fig:post_marm_openmodels_marginals}.}
  \label{fig:post_marm_openfwi_marginals}
\end{figure*}

Moving on to the training dataset, Figures~\ref{fig:post_marm_openmodels} and~\ref{fig:post_marm_openfwi} compare the inversion results obtained using diffusion models trained on the first three and the last three subsets of impedance models described above, respectively. These results reveal that, apart from the diffusion model trained on the dataset purely composed of patches from the Marmousi model, the diffusion models trained on the OpenFWI datasets tend to perform better compared to those trained on openly available velocity models (even when the Marmousi model is included in the training data). Moreover, whilst all of the standard deviations show higher values in correspondence to interfaces between layers of different impedance, those obtained from the diffusion models trained on the OpenFWI-fault and OpenFWI-nostyle datasets show generally higher values across the entire model. Using the \textit{99\% coverage ratio} metric (i.e., the percentage of acoustic impedance values of the true model that fall within the 99\% confidence interval defined from the estimated standard deviation), I conclude that the diffusion models trained on the openly available velocity datasets tend to underestimate the standard deviation of the posterior distribution. On the other hand, the diffusion models trained on the OpenFWI dataset provide a more reliable estimate of the posterior uncertainty with coverage ratio values as high as 83.6\%. However, in this case, PGDM seems to outperform DSP for some training sets whilst the opposite happens for others.

As a final metric of comparison, I compare four realizations from some of the diffusion models described in this section using the PGDM sample. The realizations produced by the diffusion models trained on the OpenFWI dataset are clearly of lower quality compared to those obtained from the diffusion models trained on the openly available velocity models. More specifically, the former acoustic impedance models present small blocky features that are not geologically realistic (e.g., Figure~\ref{fig:post_marm_pgdmreals_openfwi1}); on the other hand, the acoustic impedance models obtained from the diffusion models trained with openly available velocity models are much more continuous and show small variations that are geologically plausible (see close-ups for each realization). Based on this analysis, it is difficult to pinpoint a single best performing model as some models seem to perform better than other depending on the metric chosen to evaluate them. Finally, it is worth noting that among the 6 training datasets, the one that includes the Style-A class of the OpenFWI dataset is the one that behave the most differently from the others; in fact, the mean solution is rather blurry and the standard deviation is more homogeneous across the entire subsurface instead of being higher at the interfaces. This behavior can be justified by looking at the individual realizations (Figure~\ref{fig:post_marm_pgdmreals_openfwi}), which are also blurry, likely due to the fact that the models in the Style-A class are much smoother than any of the models coming from the other 4 classes (and from the openly available velocity models). This result clearly shows the importance of choosing a representative prior, or more specifically a training dataset that is representative of the subsurface model that one expects to invert the seismic data for.

Finally, Figures~\ref{fig:post_marm_openmodels_1dstat} and ~\ref{fig:post_marm_openfwi_1dstat} show the marginal probably density functions along 4 vertical profiles for the DPS samples obtained using the Other and OpenFWI-nostyle training datasets. As previously noted, the marginal distributions from the first model are narrower than those from the second model (also confirmed by the 99\% the coverage ratio). Similarly, Figures~\ref{fig:post_marm_openmodels_marginals} and ~\ref{fig:post_marm_openfwi_marginals} show marginal probably density functions for 4 points in the domain of interest obtained from the same two datasets using both the DPS and PGDM samplers. Overall, for a given training dataset the marginals obtained from the DPS algorithm tend to be wider than those from the PGDM algorithm, and again those obtained using the OpenFWI-nostyle training dataset are wider than those obtained using the Other training dataset. Moreover, one can notice how the distributions for points 2 and 3 are much closer to a uni-modal distribution than those from points 1 and 4; this is in agreement with the fact that points 2 and 3 are selected in the middle of two thick layers, whilst points 1 and 4 are chosen on the boundary between two layers with relatively different acoustic impedance. Such a result clearly shows that the posterior distribution estimated by means of a measurement-guided reverse diffusion model can be as complex as required by the prior and problem of interest. In this specific case, since the modeling operator is linear, using a Gaussian distribution for the prior and likelihood will inevitably lead to a Gaussian posterior. The fact that we do not estimate a Gaussian posterior here has to be interpreted as an indication that a Gaussian prior is not suitable to represent complex multi-dimensional objects like geological models and the implicit prior learned from the diffusion model is instead likely to be non-Gaussian, which leads to a non-Gaussian posterior.\\

\begin{figure*}[!htb]
\centering
  \includegraphics[width=0.99\textwidth]{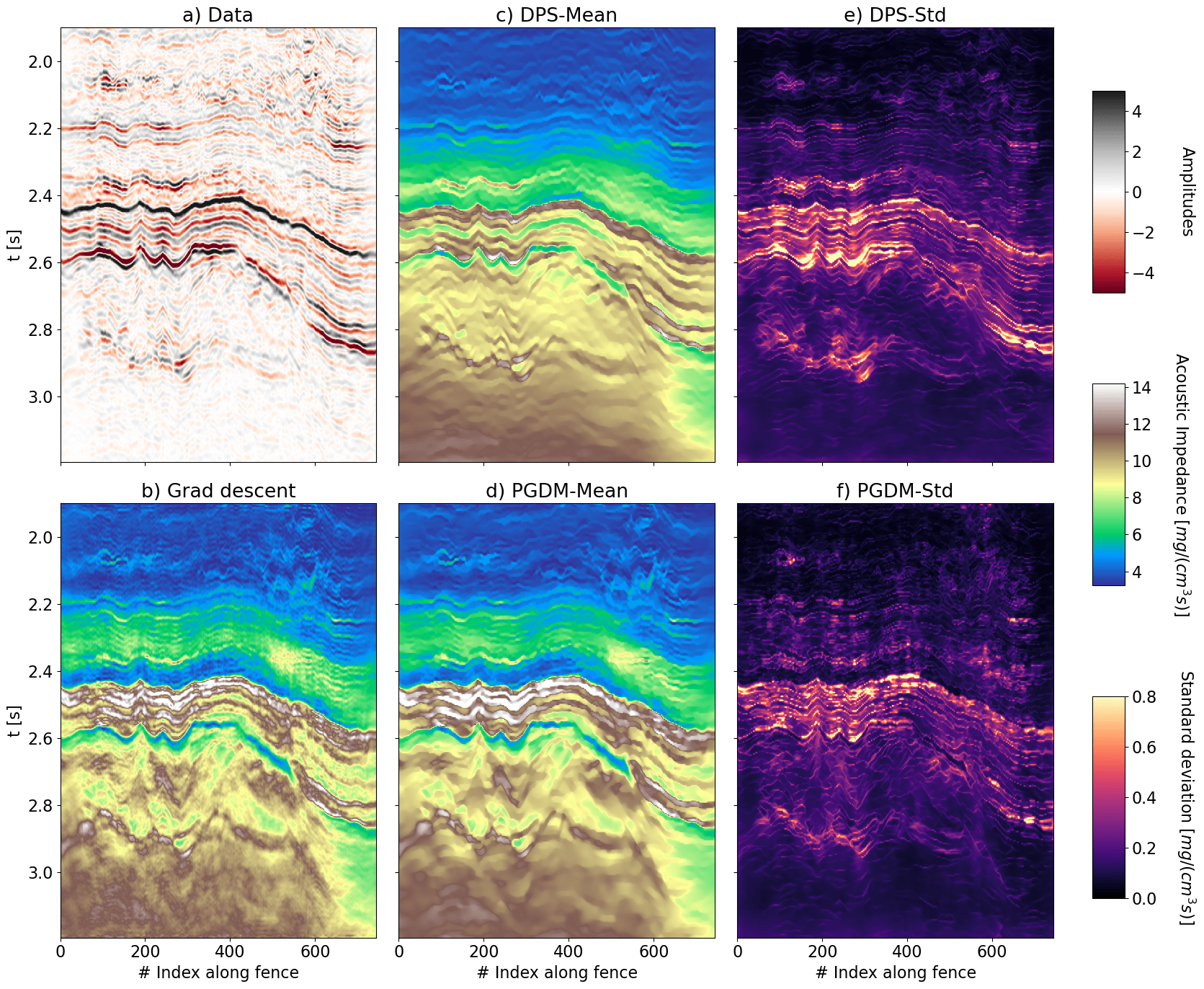}
  \caption{a) Data, b) gradient descent benchmark solution, c-e) mean and standard deviation from DPS, and d-f) mean and standard deviation from PGDM, both of them using the diffusion model trained on the Marmousi training dataset.}
  \label{fig:post_volve_comparison}
\end{figure*}

\textbf{Volve field}
Next, the diffusion models trained using the different subsets of training data are used as implicit priors in the inversion of a 2d section of the PSDM full stack dataset from the ST10010ZC11 survey from the Volve field. The section is extracted along the NO/15-9 19 BT2 well and further extended to the East of the well (Figure~\ref{fig:post_volve_comparison}a). Following~\cite{ravasi2021}, the background model is built from the root-mean-square (RMS) velocity model: RMS velocities are first converted into interval velocities and subsequently calibrated with the acoustic impedance log of the NO/15-9 19 BT2 well. More specifically, the interval velocity model is extracted along the well trajectory and regressed against the acoustic impedance log. The regression coefficients are then used to convert the entire velocity model into a background acoustic impedance model.

\begin{figure*}[!htb]
\centering
  \includegraphics[width=0.99\textwidth]{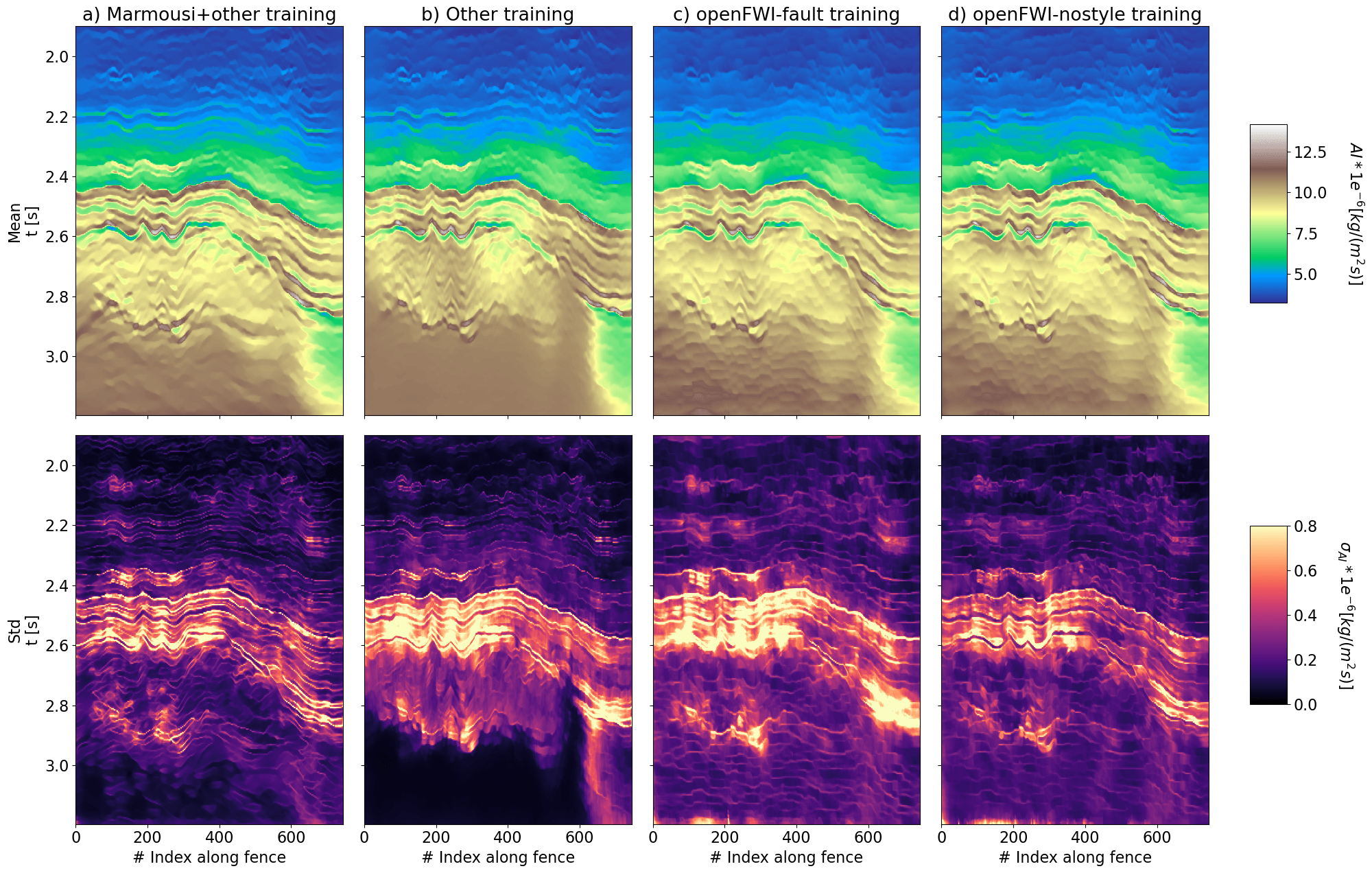}
  \caption{Mean and standard deviation for the DPS algorithm using a) Marmousi+other, b) Other, c) OpenFWI, and d) OpenFWI-fault  training datasets.}
  \label{fig:post_volve_comparison1}
\end{figure*}

\begin{figure*}[!htb]
\centering
  \includegraphics[width=0.99\textwidth]{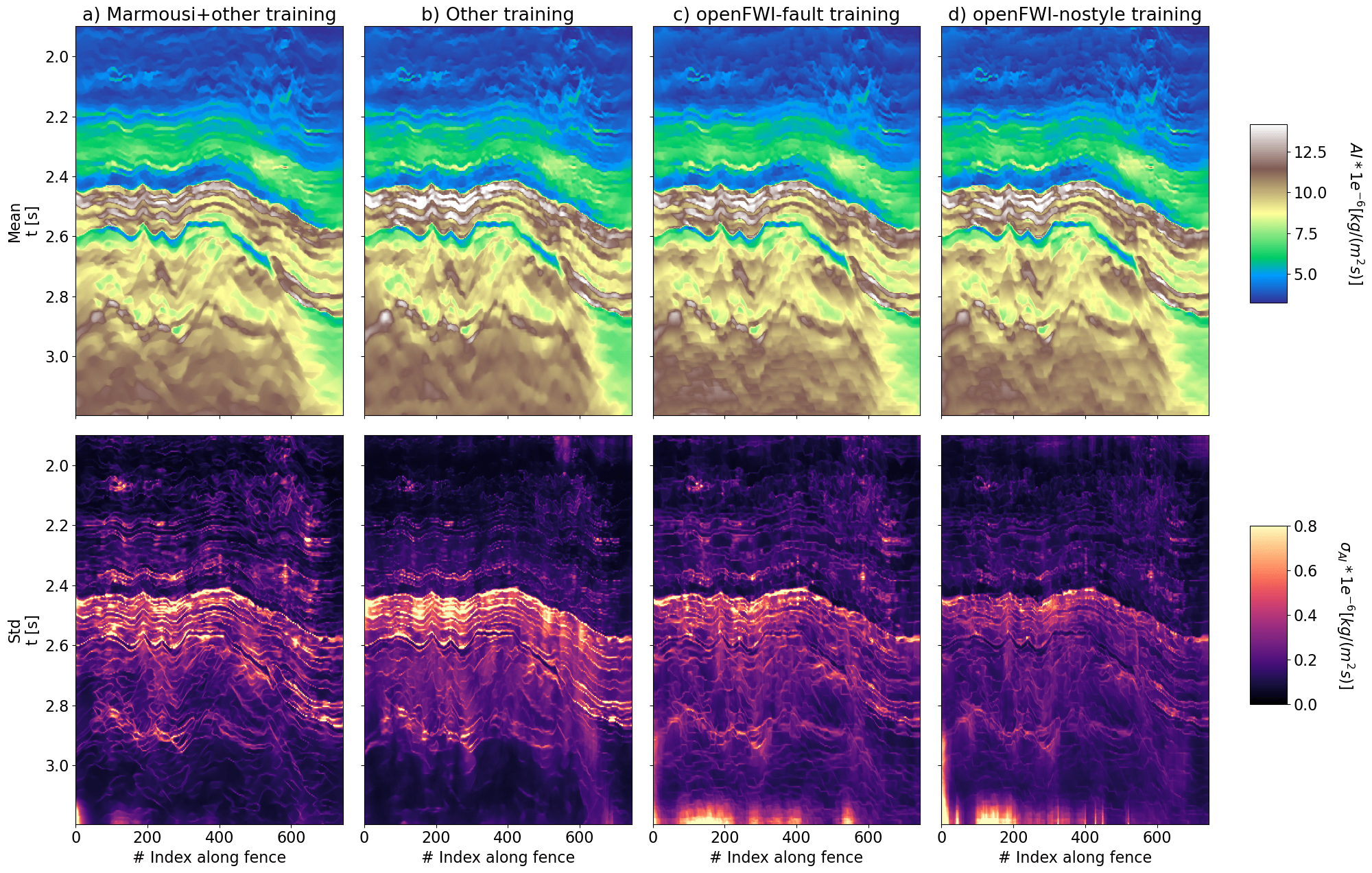}
  \caption{Mean and standard deviation for the PGDM algorithm using a) Marmousi+other, b) Other, c) OpenFWI, and d) OpenFWI-fault  training datasets.}
  \label{fig:post_volve_comparison2}
\end{figure*}

\begin{figure*}[!htb]
\centering
  \includegraphics[width=0.99\textwidth]{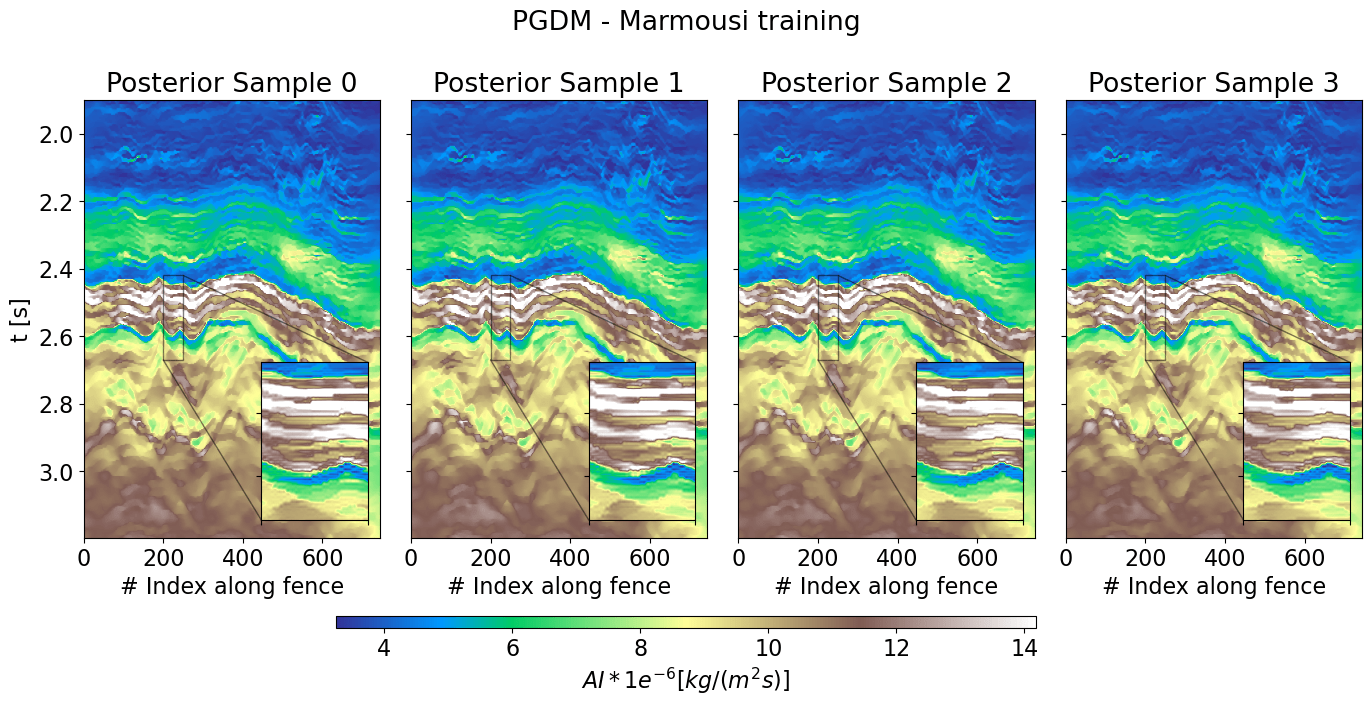}
  \caption{Realizations obtained using the diffusion model trained on the Marmousi dataset.}
\end{figure*}

\begin{figure*}[!htb]
\centering
  \includegraphics[width=0.7\textwidth]{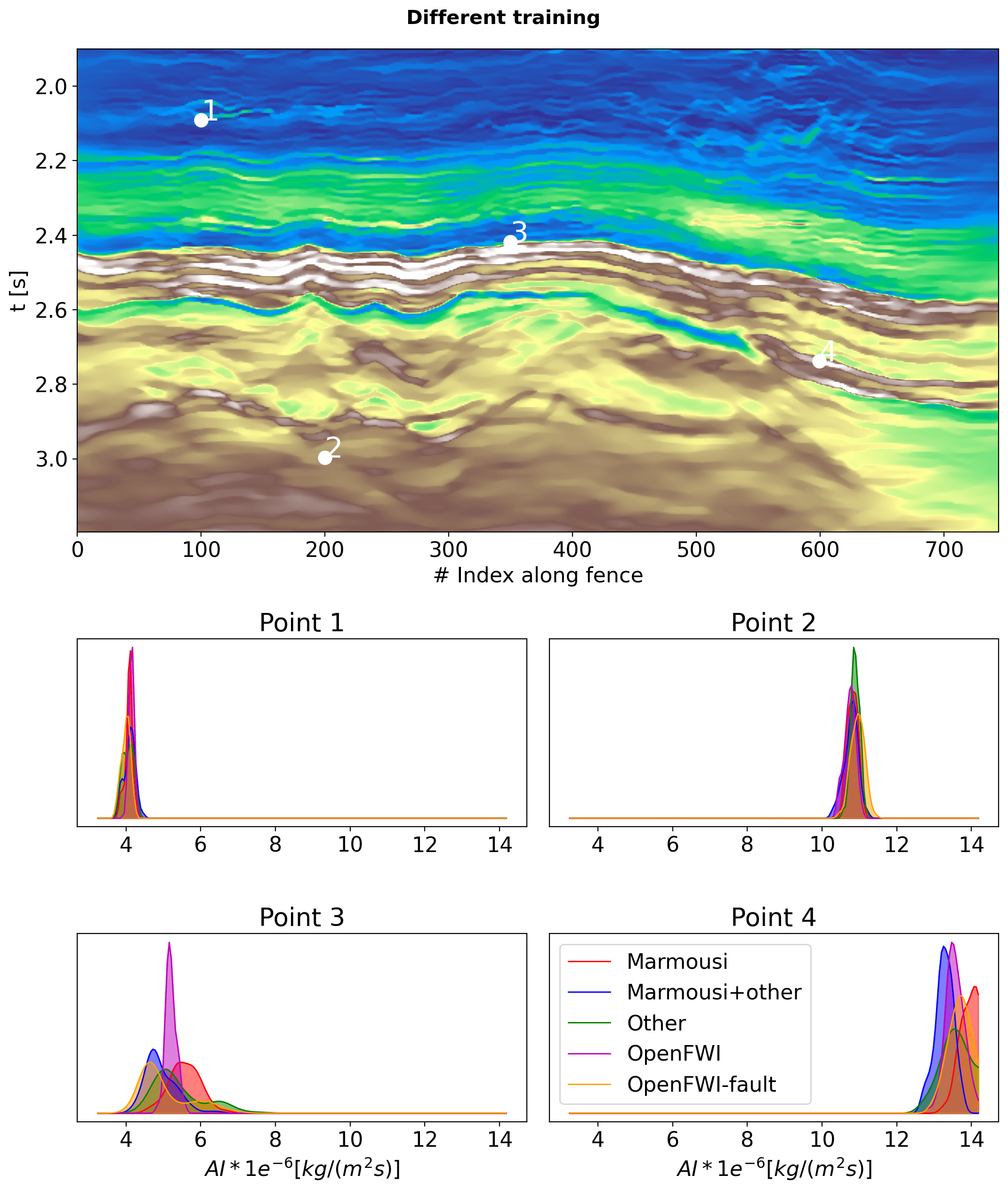}
  \caption{Marginal distributions at 4 different locations in the subsurface for different sets of realizations computed using the PGDM sampler and different training datasets as shown in the legend.}
  \label{fig:post_volve_marginalpoints}
\end{figure*}

\begin{figure*}[!htb]
\centering
  \includegraphics[width=0.99\textwidth]{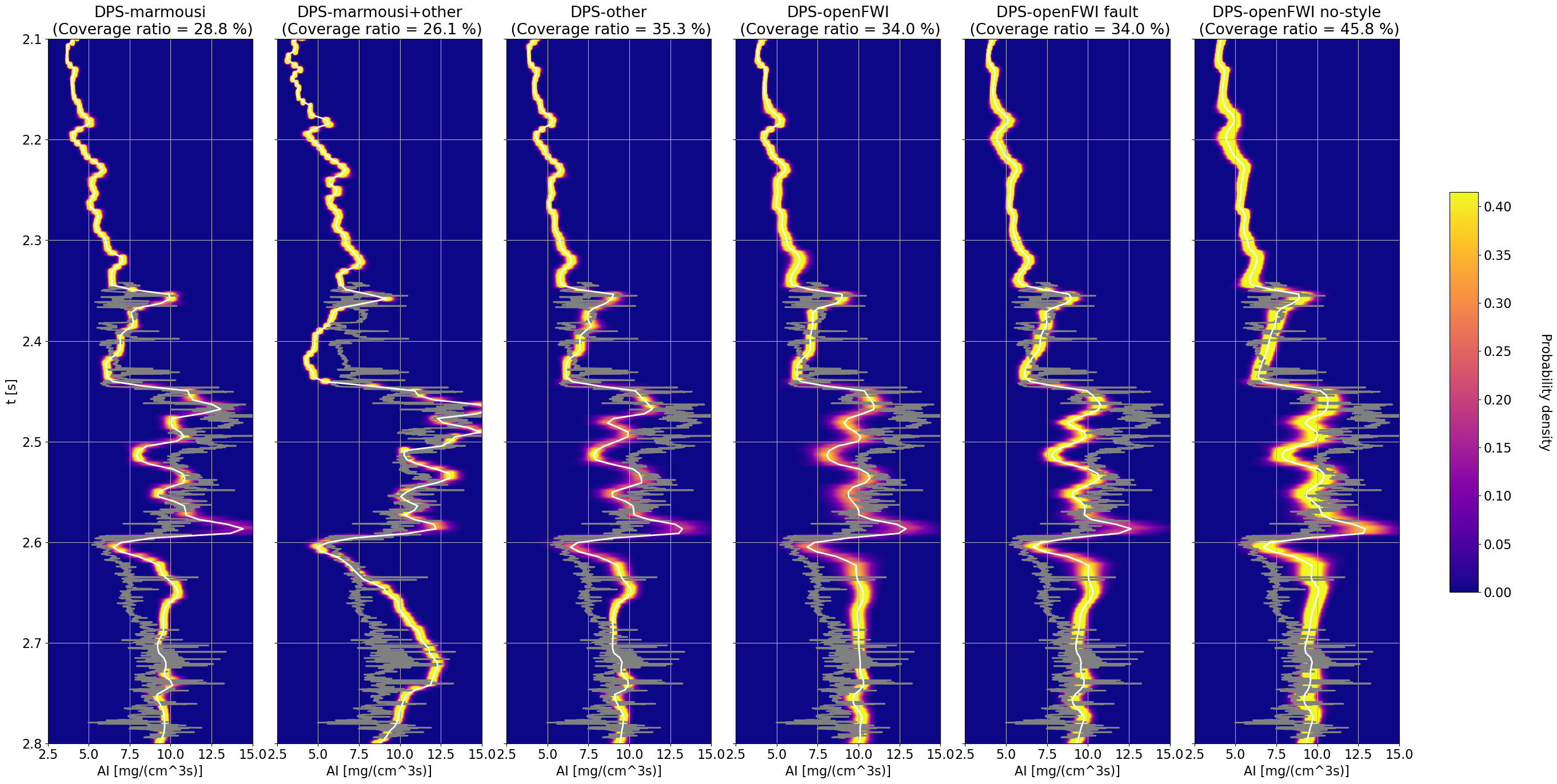}
  \caption{Marginal distribution along the well trajectory for the DPS sampler with different training data compared to the acoustic impedance well log profile.}
  \label{fig:post_volve_traceswell}
\end{figure*}

\begin{figure*}[!htb]
\centering
  \includegraphics[width=0.99\textwidth]{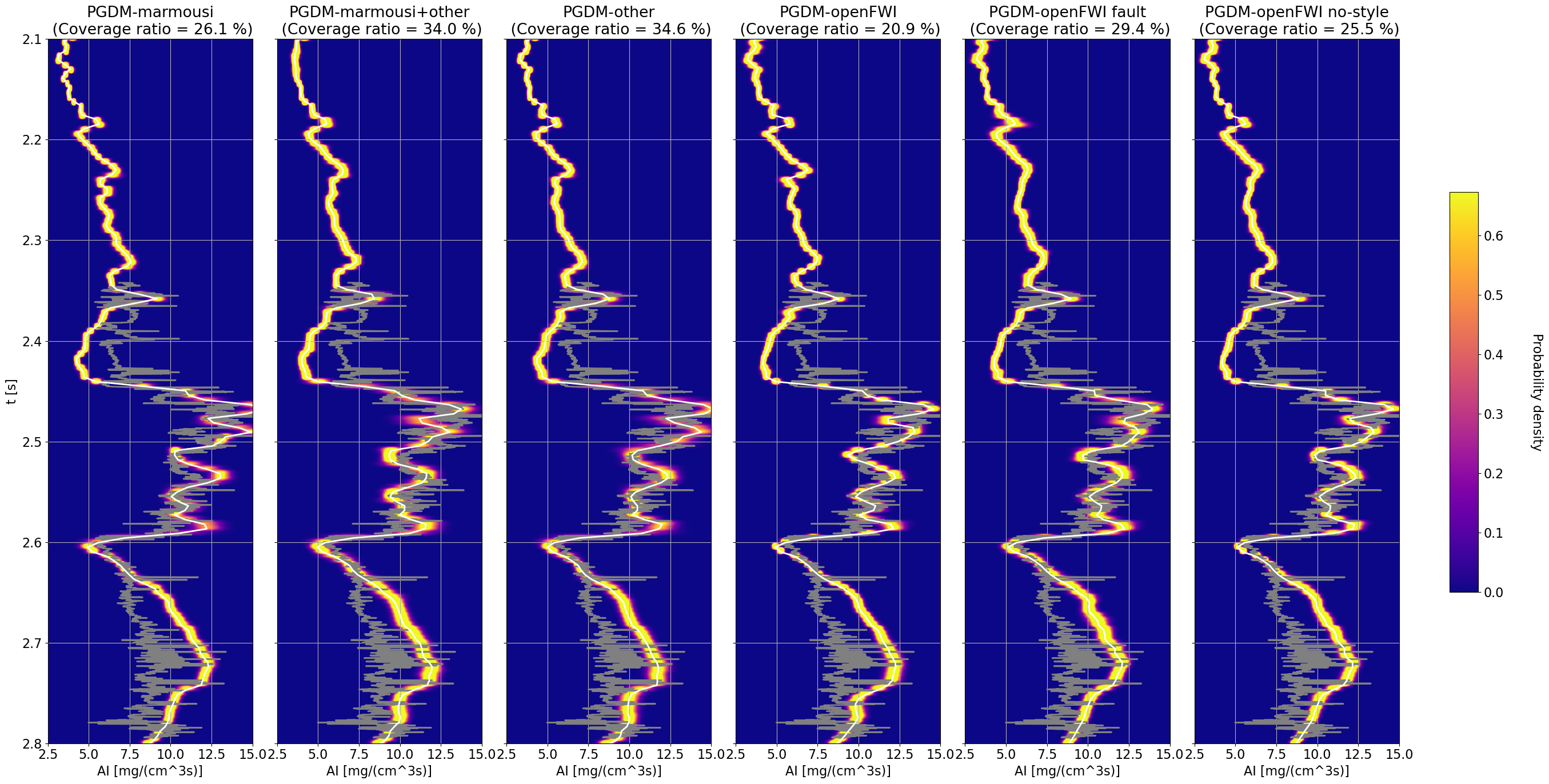}
  \caption{Marginal distribution along the well trajectory for the PGDM sampler with different training data compared to the acoustic impedance well log profile.}
  \label{fig:post_volve_traceswell1}
\end{figure*}

First, like for the synthetic example, the seismic data is inverted using a vanilla gradient descent algorithm with fixed step-size for a total number of iterations equal to 200 (Figure~\ref{fig:post_volve_comparison}b); this result provides a benchmark where only the background model and the data contribute to the solution. Whilst the inverted model is consistent with previously published results (e.g.,~\cite{ravasi2021}), a loss of continuity in the main layering structure is observed near the edges of the model and the interfaces between layers are somehow smeared. This is clearly due to the fact that no prior knowledge is not injected in the inversion process (i.e., no regularization terms are added to complement the data misfit term). Next, the DPS and PGDM algorithms are used alongside the diffusion model trained from the Marmousi training dataset: the resulting mean and standard deviation (from 100 realizations) are shown in Figures~\ref{fig:post_volve_comparison}c-d and e-f, respectively. One can immediately observed that, similar to the synthetic example, the two algorithms tend to perform slightly different from each other. More specifically, the influence of the implicit prior of the diffusion model is more evident in the DPS result compared to that of the PGDM algorithm; this is likely due to the fact that the guidance terms of latter are spectrally normalization by the pseudo-inverse operator. However, in both cases, it is possible to observe an overall increase in resolution, likely due to the fact that the training data presents blocky structures, and an improvement in the continuity of layers especially near the edges of the model. Moreover, the resulting standard deviation is also different for the two algorithms: once again, the standard deviation of the PGDM algorithm is smaller than that of the DPS algorithm. However, similar to the synthetic case, the areas of highest uncertainty tend to nicely align with the strongest reflectors in the data.

Eight additional inversion results are shown in Figures~\ref{fig:post_volve_comparison1} and ~\ref{fig:post_volve_comparison2}: the DPS algorithm is used in all experiments in Figure~\ref{fig:post_volve_comparison1}, whilst the PGDM algorithm is used in all experiments in Figure~\ref{fig:post_volve_comparison2}. These results are once again aimed at assessing the performance of the two reverse diffusion processes and the impact of the different training datasets. The resulting mean models show that the features of the training datasets are impressively translated onto the posterior models. For example, the solution obtained using the Marmousi+other dataset presents sharp and fine layers across the entire subsurface (similar to the one obtained using the diffusion model trained on the Marmousi dataset). On the other hand, the solution obtained from the Other dataset presents slightly less details especially in the deeper section, because some of the models used in the training dataset present macro-layers with less details than one can find in the Marmousi model. Similarly, the solution obtained from the OpenFWI dataset is slightly blurrier than the competitive solutions due to the inclusion of the StyleA dataset. Finally, the solution from the OpenFWI-fault and OpenFWI-nostyle datasets present very strong step-like features due to the extremely large amount of small scale features present in such dataset, although not shown here.

Similar to the synthetic case, point-wise marginals are also displayed for the acoustic impedance models estimated using the DPS sampler and various training datasets (Figure~\ref{fig:post_volve_marginalpoints}). The resulting marginal distributions. Finally, Figures~\ref{fig:post_volve_traceswell} and Figure~\ref{fig:post_volve_traceswell1} display the marginal distributions for an acoustic impedance profile along the well trajectory for the different experiments discussed in the text. In agreement with our previous findings, the models estimated using the OpenFWI training datasets tend to have a larger standard deviation and therefore 99\% coverage ratio. On the other hand, the models estimated using the Marmousi training dataset provide an overall smaller standard deviation (especially when the PGDM samplers is used).

\section{Conclusions}
Posterior sampling from an unconditionally trained diffusion model remains an elusive task due to the intractable nature of the likelihood function. In recent years, several ways to approximate the score function of the likelihood term used as guidance at each step of the reverse diffusion process have been proposed in the literature. In this work, I explore some of the most popular approaches in the context of seismic inverse problems. More specifically, I show that by leveraging the power of diffusion models to represent seismic waveforms (in seismic interpolation) and acoustic impedance models (in post-stack seismic inversion), high-quality and diverse solutions that equally explain the observed seismic data can be produced in both cases.

A fundamental pre-requisite for any diffusion model is that each sample in the training dataset must be bounded between (-1, 1); consequently, the samples generated from the unconditional reverse diffusion process are also bounded. When a measurement-guided reverse diffusion process is used to solve an inverse problem, the underlying forward problem must be therefore modified to account for such a requirement. I have proposed two practical approaches to adapt the seismic interpolation and inversion problems, making them suitable for solution through a reverse diffusion process. In both cases, I leverage the linearity of the problem and scale the measurements such that the expected solution is bounded. Moreover, for the case of seismic inversion, since the model vector is usually not already centered on zero, I take advantage of the fact that the derivative of a constant is zero and therefore solve for a perturbation model around a constant, which is added back at the end of the reverse diffusion process. I foresee that similar strategies could be also applied to other linear (and possibly nonlinear) geophysical inverse problems to render them suitable to reverse diffusion processes.

Finally, a major contribution of this work lies in the systematic analysis of two key components to the success of a measurement-guided reverse diffusion process, namely the guidance term and training dataset. When the training dataset is readily available, as it is for example the case in the seismic interpolation problem (provided that one can exploit reciprocity to find a suitable domain where either sources or receivers are well sampled), the impact of the sampling algorithm can be significant: as an example, in the synthetic example with dithered subsampling, the best performing sampler improves upon the worst performing sampler by more than 4dB. On the other hand, when the training dataset is not directly available for the problem at hand, the choice of a suitable training dataset represents the main decisive factor to the success of the reverse diffusion process. As shown in the post-stack seismic inversion example, the quality of the estimated mean and standard deviation (as well as the individual realizations) is heavily influenced by the type of acoustic impedance models selected to pre-train the diffusion model. More specifically, I show that when the velocity models from the OpenFWI dataset as used, more or less high-resolution estimates can be obtained depending on whether samples from the StyleA class are included or not in the training dataset. Moreover, whilst the SNR of the mean acoustic impedance models estimated using diffusion models trained with the OpenFWI datasets is higher than that of the acoustic impedance models estimated via diffusion models trained on other openly available velocity models, the individual realizations of the latter are shown to be much more realistic. 

In conclusion, a perhaps unexpected outcome of this extensive experimentation is that the choice of the algorithm used to guide the reverse diffusion problem has a somewhat secondary impact in the final results when compared to the choice of the training dataset used to train the diffusion model upfront. In fact, numerical evidence reveals that one could easily ‘allucinate’ the solution of a inverse problems by using a training data that contains unrealistic features (e.g., extremely small scale faulted structures).

\section{Acknowledgment}
This publication is based on work supported by the King Abdullah University of Science and Technology (KAUST). The author thanks the DeepWave sponsors for their support.

\bibliographystyle{unsrt}  
\bibliography{references}

\appendix

\section{Appendix A: Unconditional sampling with different training data}
\begin{figure*}[!htb]
\centering
  \includegraphics[width=0.99\textwidth]{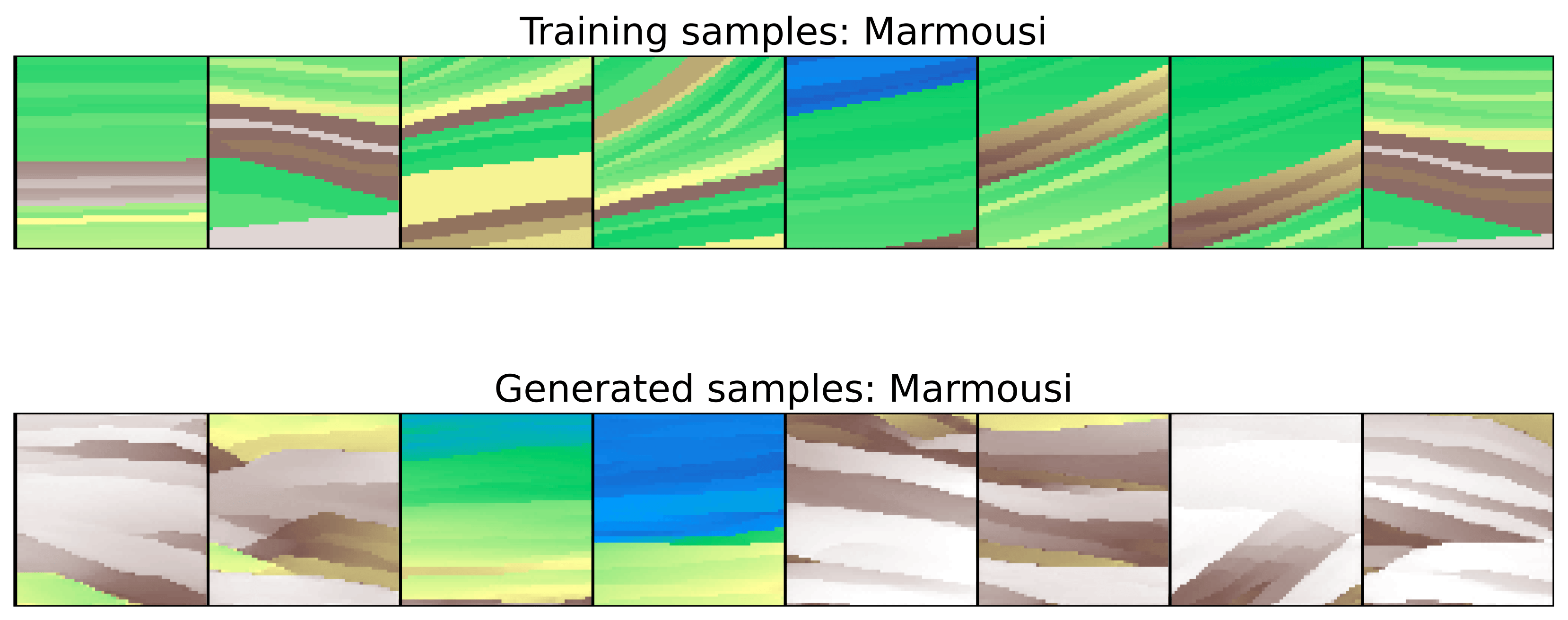}
  \caption{Top) Training samples, and bottom) unconditionally generated samples from the diffusion model trained on the Marmousi dataset.}
  \label{fig:marmousi_gen}
\end{figure*}

\begin{figure*}[!htb]
\centering
  \includegraphics[width=0.99\textwidth]{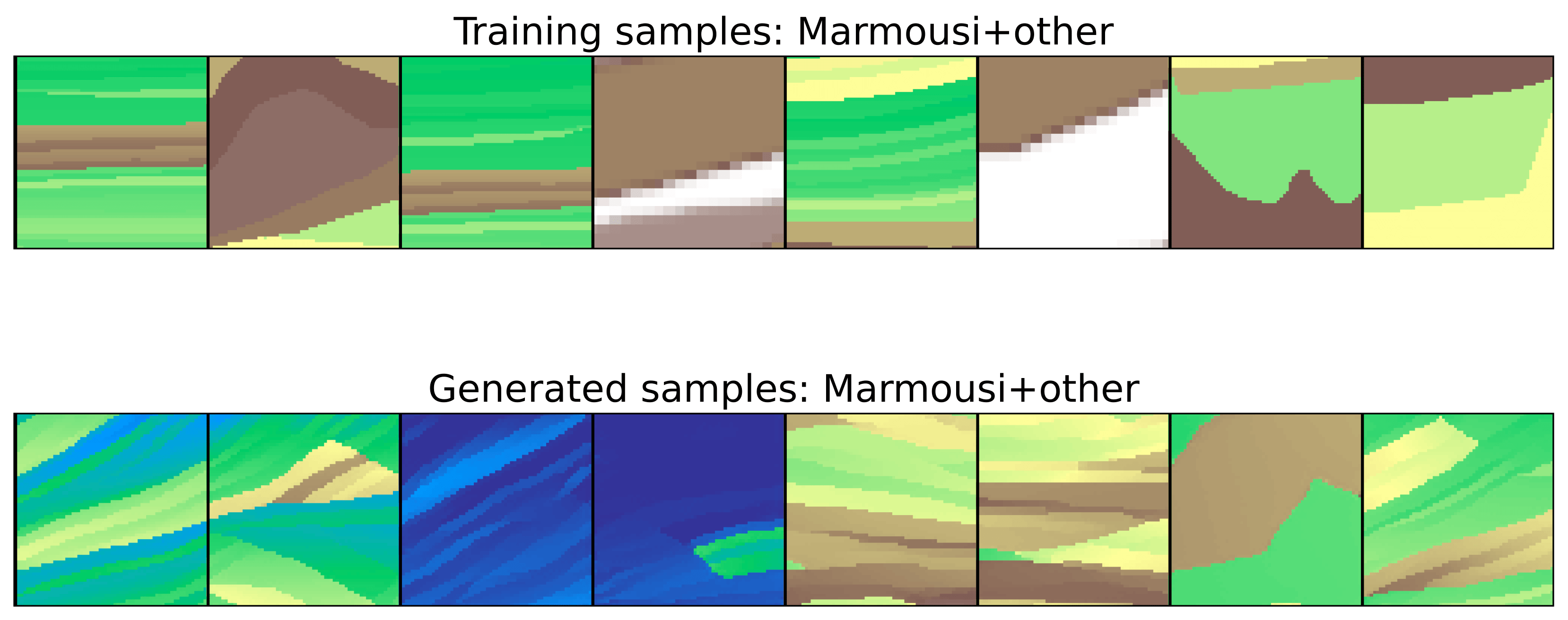}
  \caption{Top) Training samples, and bottom) unconditionally generated samples from the diffusion model trained on the Marmousi+other dataset.}
  \label{fig:marmousi_other_gen}
\end{figure*}

\begin{figure*}[!htb]
\centering
  \includegraphics[width=0.99\textwidth]{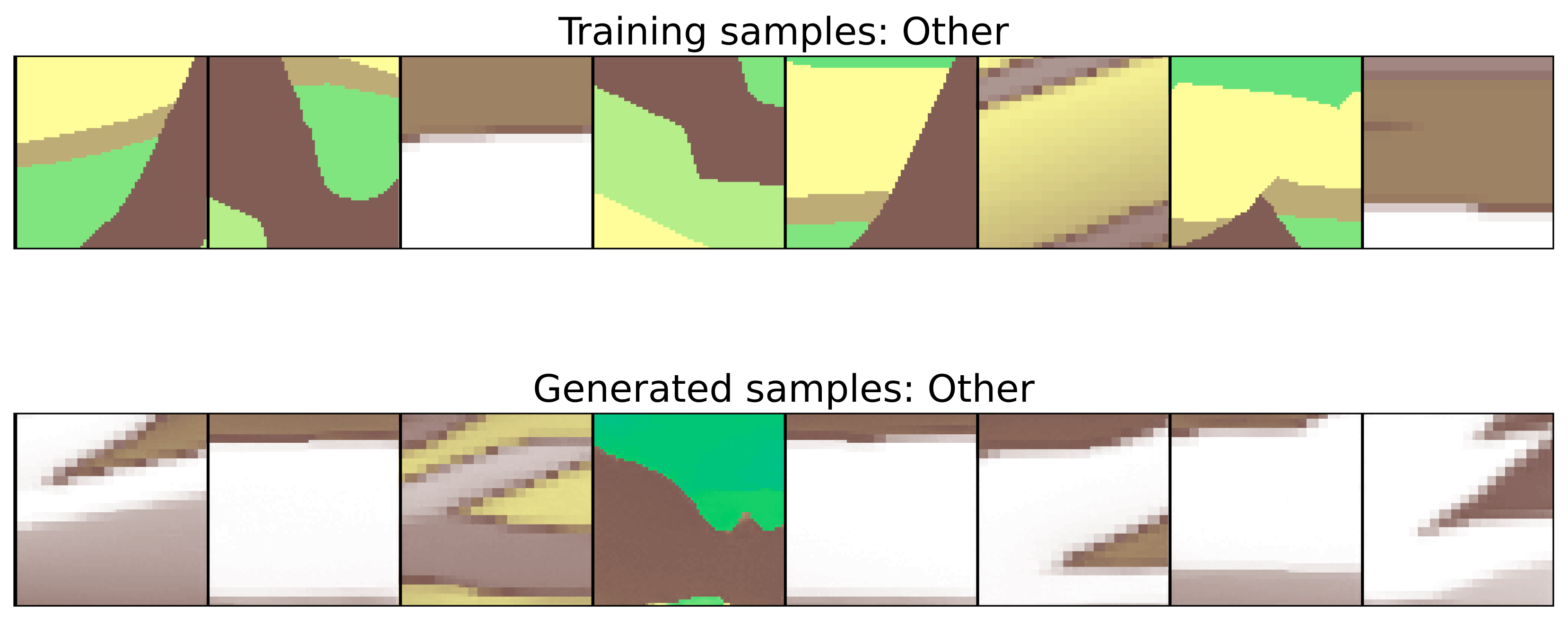}
  \caption{Top) Training samples, and bottom) unconditionally generated samples from the diffusion model trained on the Other dataset.}
  \label{fig:other_gen}
\end{figure*}

\begin{figure*}[!htb]
\centering
  \includegraphics[width=0.99\textwidth]{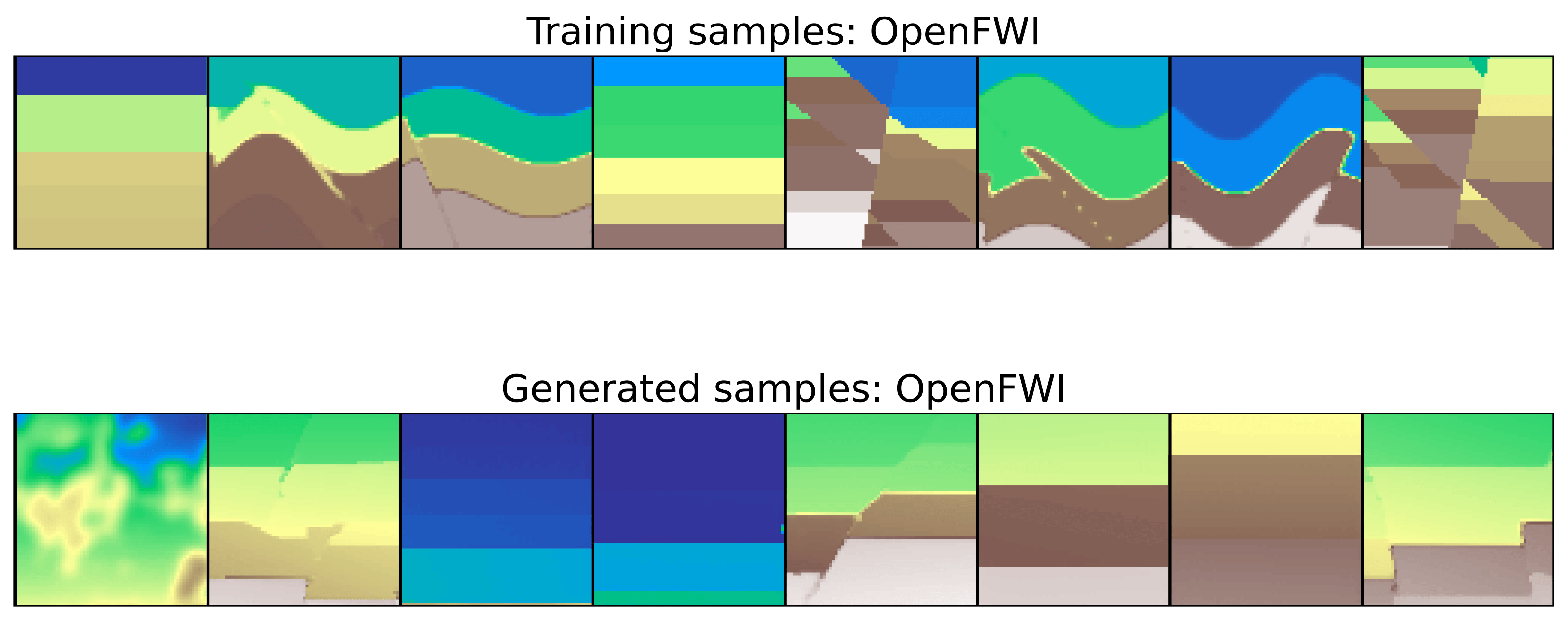}
  \caption{Top) Training samples, and bottom) unconditionally generated samples from the diffusion model trained on the OpenFWI dataset.}
  \label{fig:openfwi_gen}
\end{figure*}

\begin{figure*}[!htb]
\centering
  \includegraphics[width=0.99\textwidth]{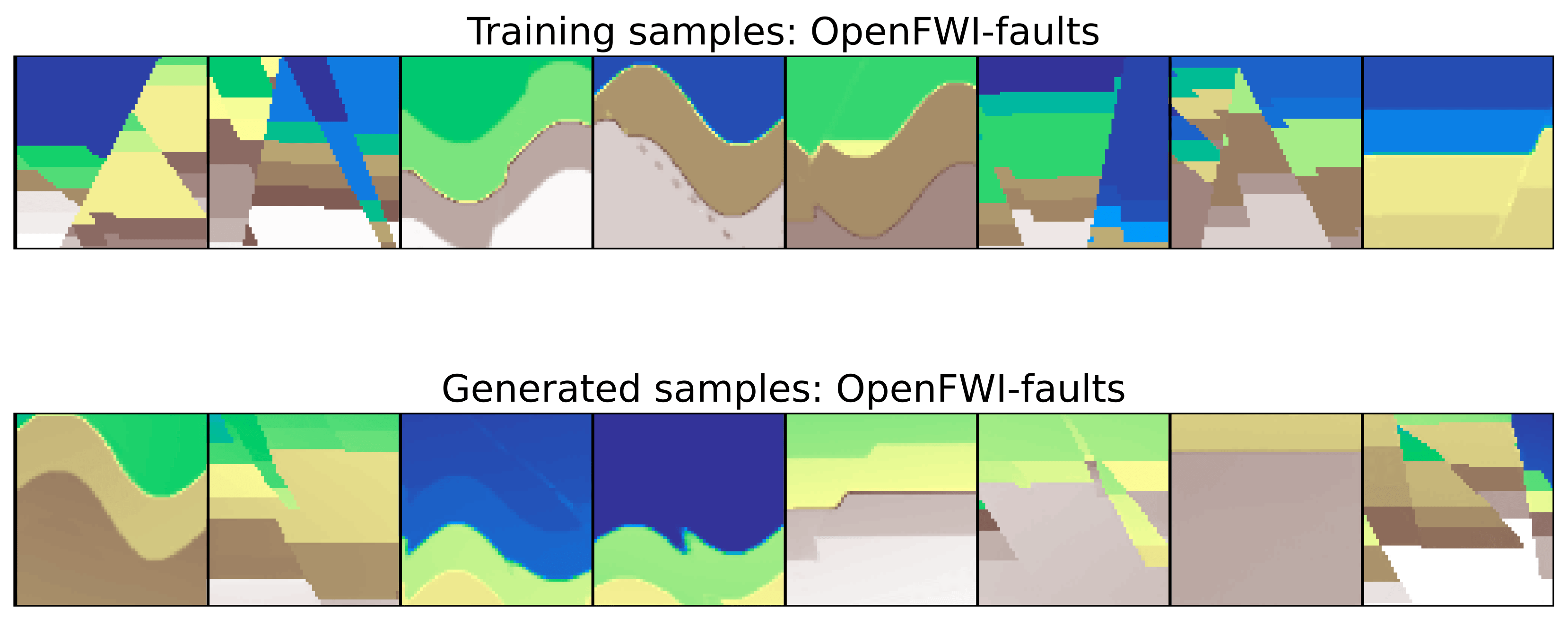}
  \caption{Top) Training samples, and bottom) unconditionally generated samples from the diffusion model trained on the OpenFWI-fault dataset.}
  \label{fig:openfwifault_gen}
\end{figure*}

\begin{figure*}[!htb]
\centering
  \includegraphics[width=0.99\textwidth]{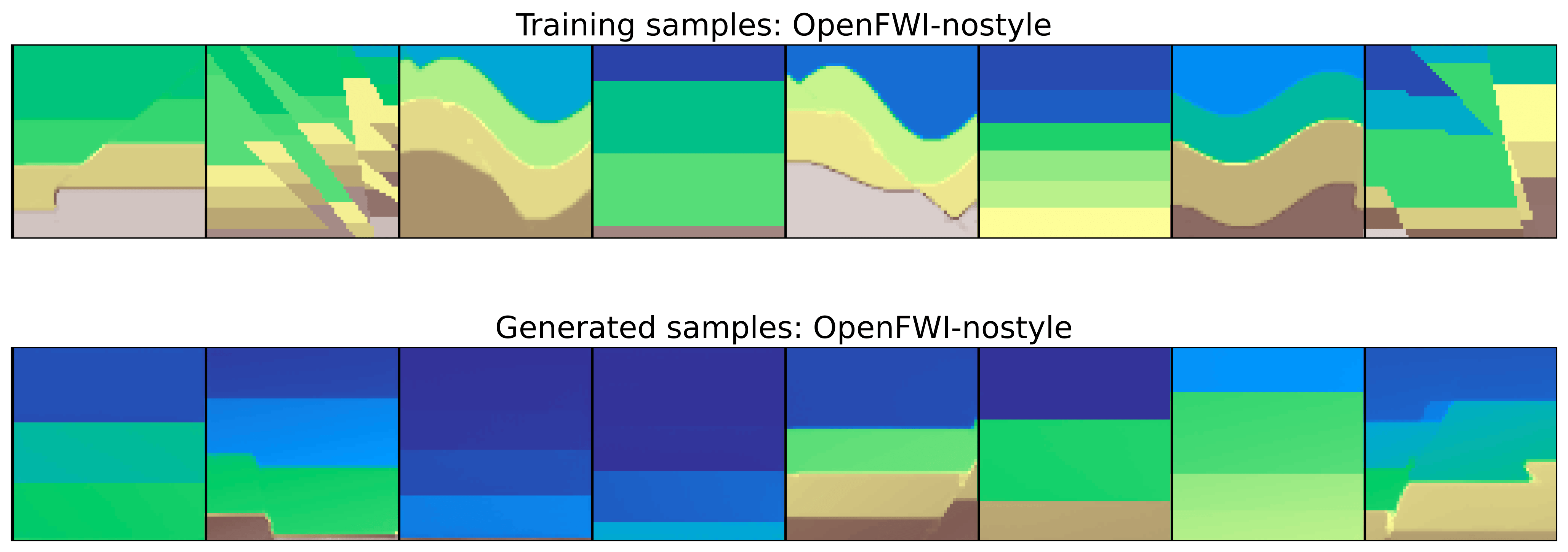}
  \caption{Top) Training samples, and bottom) unconditionally generated samples from the diffusion model trained on the OpenFWI-nostyle dataset.}
  \label{fig:openfwinostyle_gen}
\end{figure*}

In this appendix, I show eight training samples and eight generated samples for each of the training datasets described in the main body of the paper. More specifically, Figures~\ref{fig:marmousi_gen} to ~\ref{fig:other_gen} show the samples obtained by training the diffusion model on different subsets of the openly available velocity models. Similarly, Figures~\ref{fig:openfwi_gen} to ~\ref{fig:openfwinostyle_gen} show the samples obtained by training the diffusion model on different classes of the OpenFWI dataset. Finally, for all of the scenarios, two additional samples are generated using a noise realization of size $256 \times 128$ (i.e., four times as as large and twice as tall than the training patches). Note how, without any guidance from measurements, the patterns found in the smaller training patches are somehow randomly replicated across this larger domain.

\begin{figure*}[!htb]
\centering
  \includegraphics[width=0.99\textwidth]{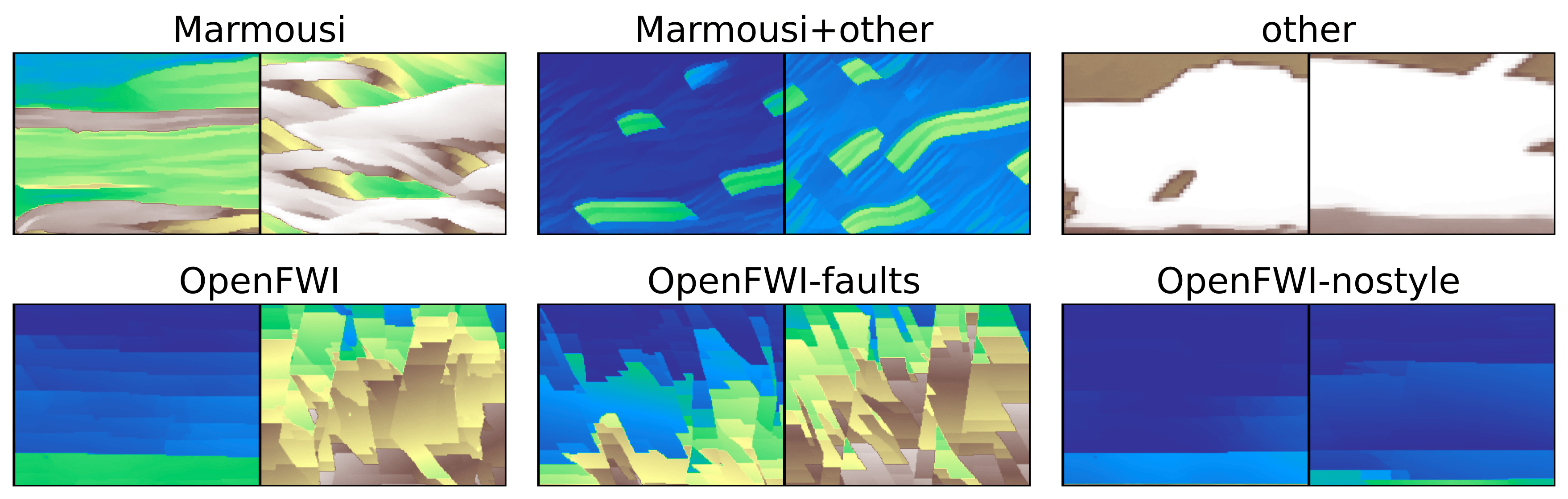}
  \caption{Samples generated on a larger domain (i.e., $256 \times 128$) by the different diffusion models trained in this work.}
  \label{fig:large_gen}
\end{figure*}

\end{document}